\documentclass{aa}
\usepackage{natbib,graphics,txfonts}

\begin{document}

\title{Galaxy evolution in groups. \\ NGC~3447/NGC~3447A: the odd couple in LGG 225} 

\author{P. Mazzei\inst{1},
A. Marino\inst{1}, 
R. Rampazzo\inst{1},  
H. Plana\inst{2},
 M. Rosado\inst{3},
\and L. Arias\inst{4}
} 

\institute{INAF Osservatorio Astronomico di Padova, Vicolo
dell'Osservatorio 5, 35122 Padova, Italy\\
\email{paola.mazzei@oapd.inaf.it}
\and
Laborat\'orio de Astrof\'isica Te\'orica e Observational, Universidade Estadual de Santa Cruz  45650-000 Ilh\'eus - Bahia Brazil
\and
Instituto de Astronom\'ia, Universidad Nacional Aut\'onoma de M\'exico,
Av. Universidad 3000, Ciudad Universitaria, C.P. 04510
\and
Universidad Iberoamericana, Department of Physics, M\'exico City, M\'exico D.F., M\'exico
%\email{lorena.arias@iuia.mx}
}
      \authorrunning{Mazzei et al.}
\titlerunning{NGC~3447/NGC~3447A: the odd couple in LGG 225}

      \date{Accepted for publication in Section 4 of  A\&A 19/10/2017 }

\abstract
{Local Group Analogs are galaxy associations dominated by few bright Spirals reminiscent of the Local Group. The NGC~3447/NGC~3447A system  is a member of the LGG 225 group, a nearby Local Group Analog. 
This system is considered a physical pair composed of an intermediate luminosity late type spiral, NGC~3447 itself, and an irregular companion,  NGC~3447A, linked by a faint, short filament of matter. 
A ring-like structure in the NGC~3447 outskirts has been emphasised by {\it GALEX} observations.}
{This work aims to contribute to the study of galaxy evolution in low density environments,  favourable habitat to highly effective encounters, shedding light on the evolution of this system. }
{We performed a  multi-$\lambda$  analysis of the surface photometry of this system to derive its spectral energy  distribution and structural properties using UV, {\it Swift}-UVOT, and  optical, SDSS images complemented with available far-IR observations.  We also characterised the  velocity field of the pair using  2D H$\alpha$ kinematical observations of the system obtained with PUMA@2.2mSPM.
All these data are used to constrain  smooth particle hydrodynamic simulations with  chemo-photometric implementation to shed light on the evolution of this system.}
{The luminosity profiles, from UV to optical wavelengths, are  all consistent with the  presence of a disc extending and including NGC~3447A. The overall velocity field does not emphasise any significant rotation pattern, rather a small velocity gradient between  NGC~3447 and NGC~3447A. Our simulation, detached from a large grid  explored to best-fit the global properties of the system, suggests that this arises from an encounter between two halos of equal mass.  }
{  NGC~3447 and NGC~3447A belong to the same halo,  NGC~3447A being a  substructure of the same disk including NGC~3447.
The halo gravitational instability, enhanced by the encounter, fuels a  long lived  instability in this dark matter dominated disk, driving the observed morphology. The NGC~3447/NGC~3447A system  may warn about a new class of  "false pairs" and the potential danger of a misunderstanding of such objects in pair surveys that could produce a severe underestimate of the total mass of the system.}


\maketitle

\section{Introduction}

In the context of group evolution, the Local Group (LG) 
offers clear examples of ongoing interaction and/or accretion episodes.
Both observations and models suggest that the galaxy evolution in the LG is highly 
dependent on the dynamic properties of the group. 
In the LG, signatures of disruption and accretion 
of small galaxies are witnessed by gaseous and stellar tidal streams 
around/onto the accreting galaxy (i.e., the Milky Way, MW hereafter), fossil relics of a  mass transfer activity.
In  recent years several relics have been found in the MW in the form of stellar
streams \citep[see e.g.][]{Odenkirchen2001,Yanny2003,Belokurov2006,
Grillmair2006}.
Concerning M~31, \citet{Block2006} found that, 
in addition to several well-known signatures of interaction 
like the warped disk and the ring at about 10\,kpc, a second  ring with a 
projected dimension of 1.5$\times$1.5 kpc is present. 
They suggest that  both ring and  warped disk have been 
originated by an head-on collision with M~32 that took place 
about 210 Myr ago \citep[see e.g.][an references therein]{Gordon2006}. 

In order to place the observed properties of the LG galaxies
in the general evolutionary framework of loose groups, we are
analysing a large sample of galaxy groups spanning a wide
range of properties, in richness of member galaxies,
velocity dispersion, morphological types and spatial
distribution with a multi-wavelength approach 
\citep{Marino2010,Marino2013,Marino2016}.
  
In this context, the case of the interacting pair NGC~3447/NGC~3447A,  alias VV 252,  (Fig.\ref{figure1}) member of the  LGG~ 225 group (Garcia, 1993), 
in the Leo cloud, is of particular interest for our project \citep{Marino2010, Mazzei2014}. This group is at a distance of %20.90$\pm$1.2\,Mpc  from Extragalactic Distance Database \citep{Tully2009}, or 
 about 15\,Mpc  \citep{Marino2010, Mazzei2014}.
NGC~3447 and NGC~3447A are considered members of a  pair composed by a
late barred spiral galaxy, NGC~3447, and an Irregular,  smaller and fainter companion,  1.2 mag fainter in the  B-band following the NED database. 
%The primary galaxy, NGC~3447, shows a UV-bright ring at a radius of $\sim$ 110$\arcsec$. {\bf 
%The ring and the  UV morphology of the system are reminiscent of an ongoing head-on collision between NGC~3447 and NGC~3447A .
 The morphology details  change with the wavelength, highlighting different
 stellar populations.   Both optical and UV images show  two distinct
 components, NGC~3447 and NGC~3447A, connected by a sort of filament. 
 UV-bands map out  the star forming complexes whose
 shape in the outskirt of NGC~3447, at a radius of $\sim$ 110$\arcsec$,
  remind a ring. So, the global morphology suggests, as a
 working hypothesis, an ongoing head-on encounter.   

Therefore, the study of  this  nearby system allows
us to investigate in great detail the effect of the interactions/collisions  on the evolution of the intermediate/low luminosity galaxies in LG Analog (LGA) environments.
This work aims to contribute to the on-going studies of galaxy evolution
in low density environments which are favourable habitat to highly effective encounters.
The paper is organised as follows. In Section 2 we describe observations and reduction of  our Fabry-Perot 2D H$\alpha$ kinematical data,  UV  {\it Swift}-UVOT, and  Sloan Digital Sky Survey (SDSS) optical images. Section 3 highlights our photometric and kinematical results which are used to constrain a large grid of smooth particle 
hydro-dynaminc (SPH)  simulations with chemo-photometric implementation (SPH-CPI simulations, hereafter). The  simulation which best fits the global properties we derived, i.e. the global SED extended over four order of magnitude in wavelength, the absolute magnitude, the morphology,  and the velocity field,   provides us with a consistent evolutionary framework of the formation and evolution of the whole system, NGC3347A/NGC~3447. These  points are  expanded in Section 4. In the Appendix we review and upgrade our knowledge of the environment of this system. Discussion and conclusions are presented in Section 5 .

\section{Observations and data reduction}\label{sec:obsred}
\subsection{\it Swift-{\tt UVOT} UV observations} \label{subsec:UVOT}

{\tt UVOT} is a 30~cm telescope on the {\it Swift} platform operating
both in imaging and spectroscopic modes \citep{Roming05}. 
UV images of  the system NGC 3447/NGC 3447A come  from the {\it  Swift} archive \footnote{www.asdc.asi.it/mmia/index.php?mission=swiftmastr}
in the three available filters, $W2$ ($\lambda_0 \ 2030$), $M2$ ($\lambda_0 \ 2231$), 
$W1$ ($\lambda_0 \ 2634$). %, $U$ ($\lambda_0 \ 3501$), $B$ ($\lambda_0 \ 4329$), $V$ ($\lambda_0 \ 5402$). 
Description of the filters, PSFs (FWHM 2\farcs92 for $W2$, 2\farcs45 for $M2$,
2\farcs37 for $W1$), %2\farcs37 for $U$, 2\farcs19 for $B$, 2\farcs18 for $V$),  and 
and calibrations are discussed in \citet{Breeveld10,Breeveld11}. 

{\tt UVOT} data obtained in imaging mode with a 2$\times$2 binning, result  in 1\farcs004/pixel, %Images were processed using the procedure described in {\tt http://www.swift.ac.uk/analysis/uvot/}.
%We combined all the images taken in the same filter for each galaxy 
%in a single image using {\tt UVOTSUM} to improve the S/N and to enhance the visibility 
%of NUV features of low surface brightness. The final data set of the  $V$, $B$, $U$, $W1$, $M2$, $W2$
and total exposure times reported in Table~\ref{exptimeUV}. 
We used the  photometric zero-points  provided by \citet{Breeveld11} to convert
{\tt UVOT} count rates into the AB magnitude system \citep{Oke74}: 
{\it zp$_{W2}$} = 19.11$\pm$0.03, {\it zp$_{M2}$} = 18.54$\pm$0.03, 
{\it zp$_{W1}$} = 18.95$\pm$0.03. % {\it zp$_U$} = 19.36$\pm$0.02,  {\it zp$_B$} = 18.98$\pm$0.02, and {\it zp$_V$} = 17.88$\pm$0.01.

{\tt UVOT} is a photon-counting instrument and is therefore subject to 
coincidence loss when the throughput is high, whether this is
due to  background or source counts,  which may result in an undercounting of the flux. 
This effect is a function of brightness of the source and affects
the linearity of the detector.    %Moreover, this produces  a misplacement of the source position due to the centroiding algorithm.  
Without binning, 
 count rates less than 0.007 counts~s$^{-1}$~pixel$^{-1}$ 
are affected by at most 1\%  and count rate less than ~0.1  counts~s$^{-1}$~pixel$^{-1}$ 
by at most 12\% due to coincidence loss  \citep[their Figure 6]{Breeveld11}. The previous values  become 0.028  and  ~0.4
counts~s$^{-1}$~pixel$^{-1}$  respectively, accounting for  2$\times$2 binning.
We checked  for coincidence loss  all our  images finding that these are not affected by the problem.

Our images % taken from the {\it  Swift} archive (www.asdc.asi.it/mmia/index.php?mission=swiftmastr) 
are  the sum of several (dithered and rotated) frames. 
This sum tends to smooth out large-scale inhomogeneities in the final frame.
 Our target covers a limited portion of the  17\arcmin$\times$17\arcmin
 of the {\tt UVOT} field of view and the background can be well evaluated around it.

\begin{figure}
\centering
\includegraphics[width=10.cm]{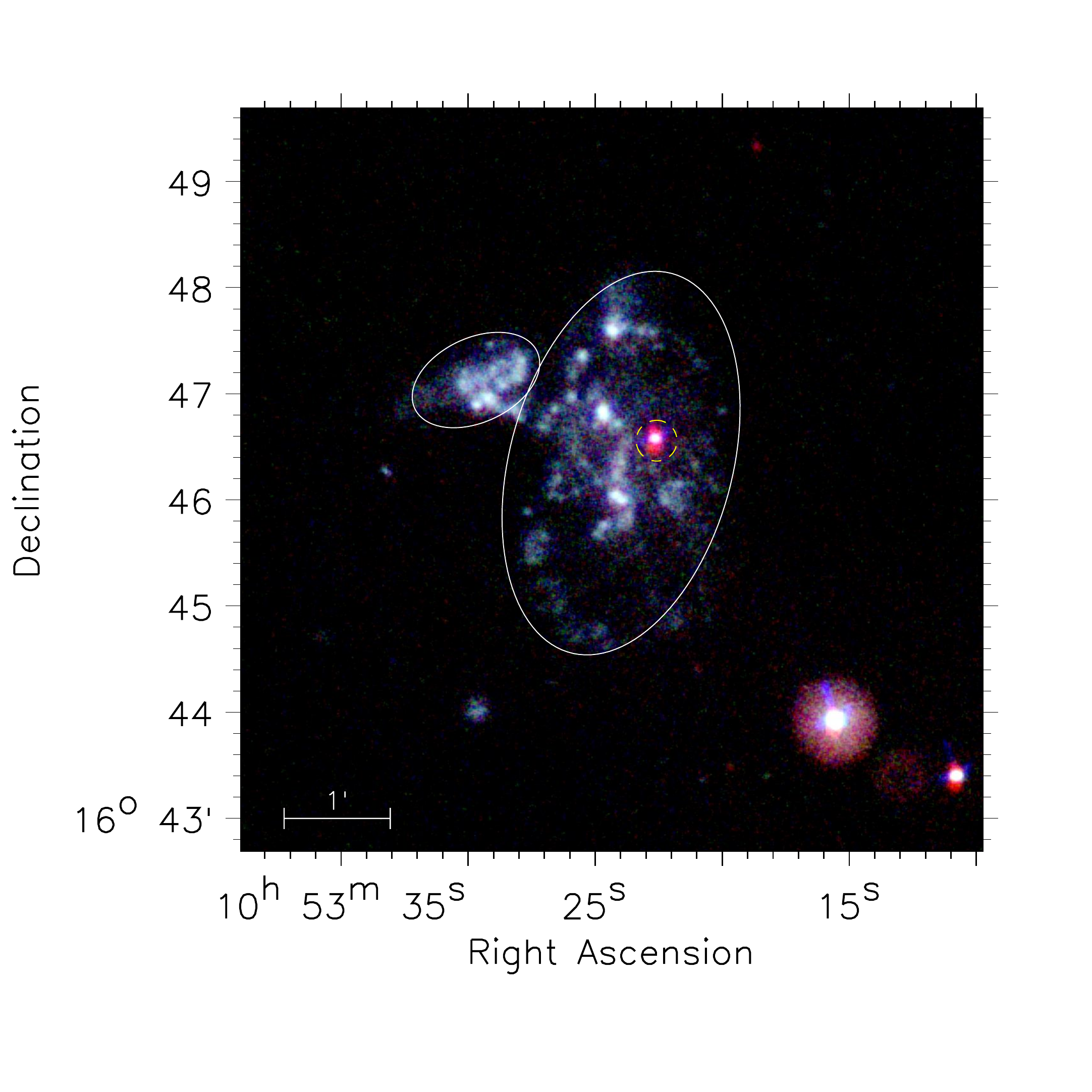}
\caption{  Color composite UV image  in the {\it Swift}-{\tt UVOT}  filters,
 $W2$ blue, $M2$ green and $W1$ red,   of the pair system, NGC~3447 and NGC~3447A (white ellipses). %;  the field of view is 7$\arcmin\times$ 7$\arcmin$, North on the top and East to the left. 
 The emission within the dashed circle  corresponds to the SNa 2012th \citep{Yamanaka14}, masked in our photometric analysis (Sect.\ref{subsec:UVOT}).}
\label{figure1}
\end{figure}

Surface photometry has been performed as  described in \citet{Rampazzo2017}, i.e.  using the 
{\tt ELLIPSE} \citep{Jedr87} fitting routine in the {\tt STSDAS} 
package of {\tt IRAF}, increasing the size of the apertures logarithmically. 
Foreground and background objects outside the galaxy body as well as the  SN on our target (see Fig.\ref{figure1} ) have been masked.
To secure a reliable background estimate, we performed the measurement well beyond the galaxy emission. {\tt ELLIPSE} provides the semi-major axis lengths ($a$), the surface brightness ($\mu$), the ellipticity ($\epsilon$), the position angle (PA), 
and the isophotal shape ($a_4/a$) 
%This  term, labeled $B_4$ in the {\tt ELLIPSE} table,  provides the deviation from the elliptical shape, which is parametrized by the fourth cosine  coefficient of the Fourier expansion of the residuals of the fitting procedure. 
%The sign, the absolute value, and the behavior of $a_4/a$ are indicative of 
%the boxiness ($a_4/a<0)$ or diskiness ($a_4/a$>0)  of the isophotes 
\citep{Bender89,Capaccioli90,Governato93}. \\
\indent
We derived total apparent AB
magnitudes by integrating the surface brightness within elliptical isophotes.
 Errors were estimated by propagating the
statistical errors on the  isophotal intensity provided by {\tt ELLIPSE}.    
%These profiles (Fig.\ref{SDSSprof}, left panel)  are used to perform the structural analysis of
%our system. 
Our UV  total magnitudes, reported in Table~\ref{exptimeUV}, are corrected for Galactic extinction following \citet{Roming09}. \\

%-----------------------------------figure1----------
%-------------------------- end figure1 -------------------------------------

%-------------------- Table 1 --------------------------------------- 
\begin{table*}
\center
\caption{ {\it Swift}-{\tt UVOT} images: exposure times and integrated  magnitudes in the [AB] system}
\begin{tabular}{ccccccc}
\hline
Filter & $W2$ & $M2$ & W1   \\
\hline
\hline
%Product &00032662004 & 00032662017 & 00032662003 \\
Exp. times  [sec]  &1123  & 1130  & 1657 \\
mag [AB] & 13.95$\pm$0.10& 13.93$\pm$0.11& 13.83$\pm$0.10&\\
\hline
\end{tabular}
%\footnotesize{The product 00085541005 (2105) does not include the SN and in the M2 filter provides 1474 sec (longer than the exposure selected,  however M2 magnitude is the same as in this Table.

\footnotesize{Magnitudes  corrected for galactic extinction following \citet{Roming09}.
}
\label{exptimeUV}
\end{table*}
%-------------------- end Table 1 ----------------------------------

\subsection{SDSS data }
\begin{figure*}
\centering
\includegraphics[width=6.5cm]{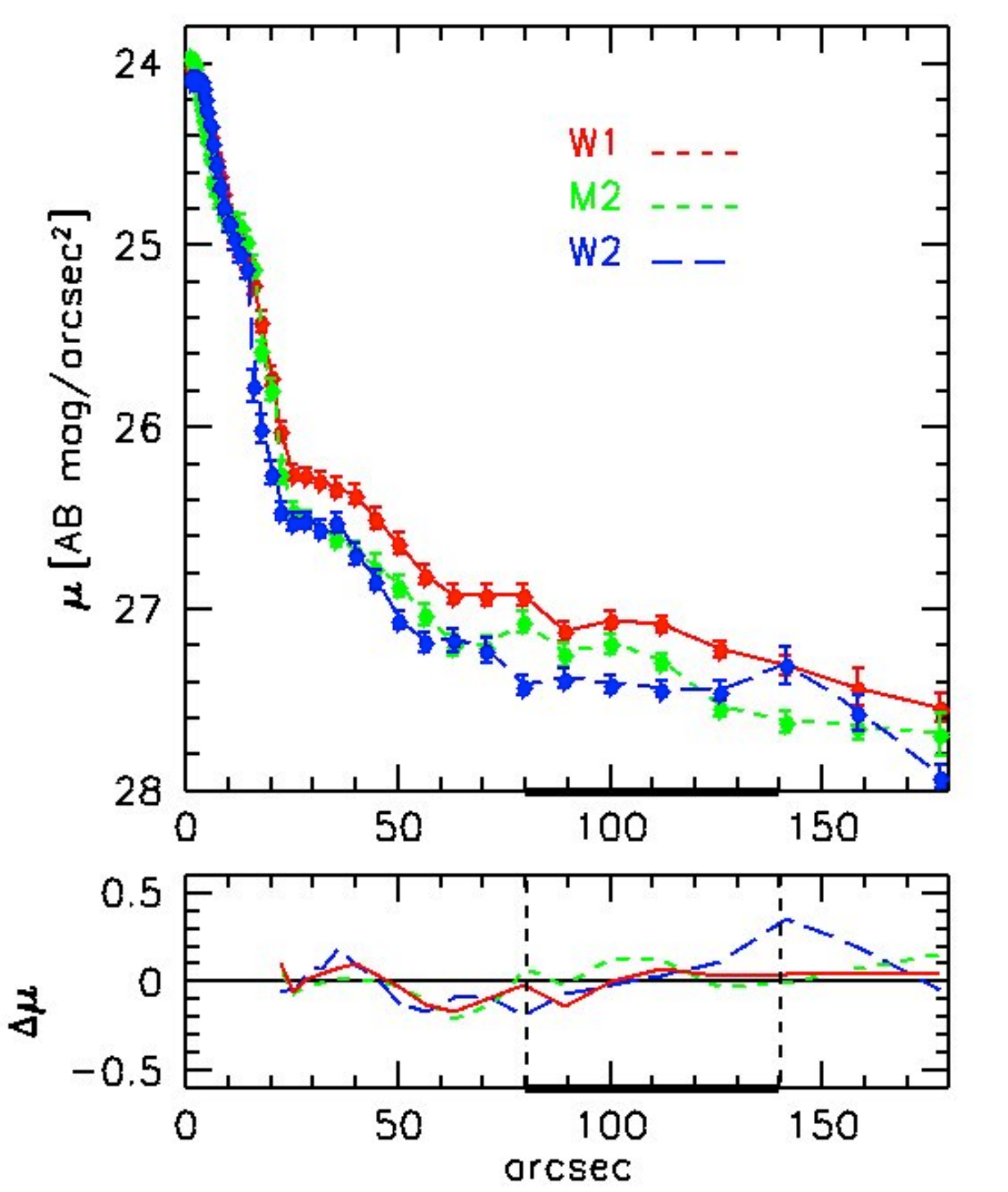}
\includegraphics[width=6.5cm]{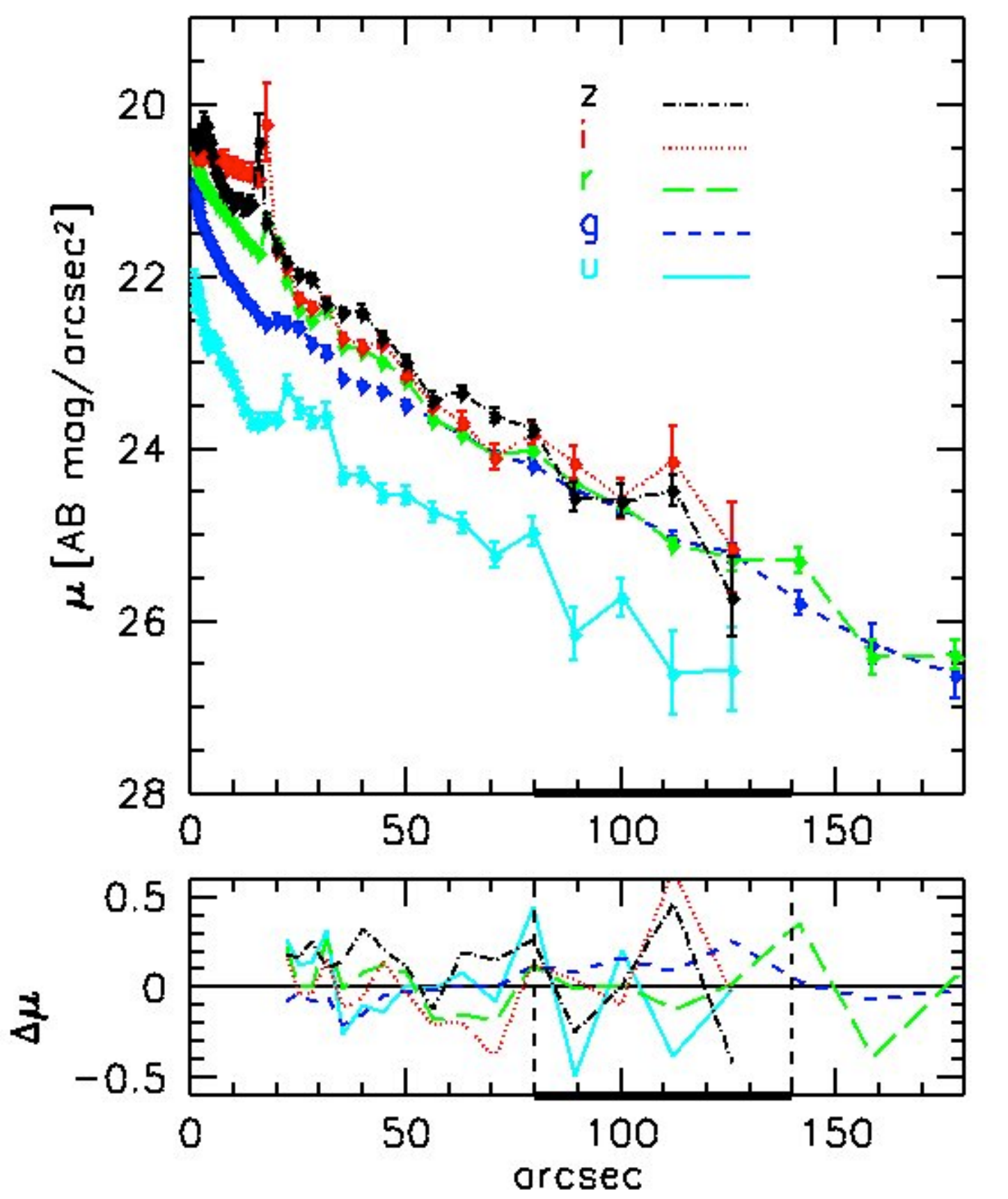}
\caption{ 
The surface brightness profiles in the {\tt UVOT} $W2$, $M2$ and $W1$  bands (left) and SDSS, $u$, $g$, $r$, $i$, and $z$ (right). All the profiles include  NGC~3447A,
extending from 80\arcsec to 140\arcsec,   as emphasized (bold) in the figure.
The  S\'ersic fit  of all the profiles  provides n=1.4. The model has been subtracted from the original profile and the residuals are shown in the  bottom panels.
The spike arising at 18\arcsec,  from r to z-bands, corresponds to the bright H$\alpha$ region North of the bar.
Profiles are  not corrected for galactic extinction and truncated when the error bars exceeds 0.5\,mag.  
}
\label{SDSSprof}
\end{figure*}

We used u, g, r, i, and z images mosaiced by \citet{Knapen2014a}. These are taken from
SDSS DR7 and SDSS-III DR8 science archives. These homogenised, background subtracted and calibrated images  cover  at least $3 \times D_{25}$, where $D_{25}$ is
the galaxy diameter at 25 mag\,arcsec$^{-2}$ in each of the SDSS filters.  The pixel size is 0\farcs396\,px$^{-1}$;  the zero points   are provided in the keyword {\it magzpt} of the header of each released image. The product we used (PGC0032694) provides a 10\farcm1 FoV.
The AB integrated magnitudes we derived are reported in Table \ref{SDSS} accounting for galactic extinction correction. Figure \ref{SDSSprof} (right panel) shows the surface brightness profiles. In the same table we also report the corresponding AB magnitudes from Table 4 of \citet{Marino2010}. All the magnitudes are corrected for galactic extinction from NED.
%-------------------- Table 2 --------------------------------------- 
\begin{table*}
\centering
\caption{ SDSS integrated  magnitudes in the [AB] system}
\begin{tabular}{ccccccc}
\hline
Filter & $u$ & $g$ & $r$ & $i$ & $z$ &  \\
\hline
\hline
%Product &PGC0032694  &PGC0032694   &  PGC0032694 &PGC0032694   & PGC0032694   & \\
%Exp. times  [sec]  &162.7  & 161.7  & 161.7 & 161.7 & 161.7 \\
mag [AB] & 13.55$\pm$0.16& 12.57$\pm$0.05& 12.30$\pm$0.05&12.17$\pm$0.08&11.50$\pm$0.23\\
mag [AB]$^1$ & 13.94$\pm$0.13& 12.91$\pm$0.05& 12.64$\pm$0.06&12.55$\pm$0.08&11.89$\pm$0.14\\
\hline
\end{tabular}

\footnotesize{$^1$ from Table 4 of \citet{Marino2010}.\\
}
\label{SDSS}
\end{table*}
%-------------------- end Table 2 ----------------------------------

\subsection{2D H$\alpha$ kinematical observations and data reduction}
\label{subsec:Fabry-Perot observations}
%\subsection{Fabry-Perot observations}
Observations \footnote{All Fabry-Perot data (cubes and moment maps) are available at the CDS via anonymous ftp to cdsarc.u-strasbg.fr (130.79.128.5)}
were carried out  with the PUMA Fabry-Perot (FP, hereafter) interferometer at the 2.1 m telescope of the Observatorio Astron\'omico Nacional at San Pedro M\'artir (OAN-SPM) Baja California, M\'exico,
on March 13$^{th}$ 2013.
The PUMA  instrument \citep{Rosado95} 
is formed by a Queensgate ET50 etalon in the pupil of a focal reducer with the Marconi2 2K$\times$2K CCD detector as receptor. Marconi2 has a physical pixel size of 13\,$\mu$m  and a scale on the sky of 0\farcs33/px.  We  performed 4$\times$4 spatial binning giving a pixel equivalent of 1\farcs3 on the sky.
Table \ref{FP} provides the journal of observations and characteristics of the interferometer we have used.\\
\indent
 The PUMA data have been reduced using the {\tt Adhocw} software package\footnote{Available at https://cesam.lam.fr/fabryP\'erot/index/softwares} and {\tt Computeverything}\footnote{Available at http://www.astro.umontreal.ca/~odaigle/reduction/} by \citet{Daigle06}.
The first step, before the correction phase, is to perform the standard CCD data reduction by applying bias and flat-field corrections to each frame of the data cube. 
The object acquisition procedure and the  data reduction have been  described by \citet{Amram96} and \citet{Daigle06}; see also \citet{Sanchez2015} for the specific case  of data cubes acquisition directly related to the PUMA instrument.\\
\indent
 In addition, a channel to channel transparency correction was done using several zones around the object.
Phase map has been built by scanning  the narrow H$\alpha$ 6563\AA\, line under the same observing conditions and phase corrected cubes have been performed using the {\tt Adhocw} package.
Velocities are measured relative to the derived systemic velocity of 1079$\pm$10\,km\,s$^{-1}$ with an accuracy of 10\%. The signal measured along the scanning sequence was separated into two parts: (i) an almost constant level produced by the continuum light in a 89\AA\, bandpass around H$\alpha$ (continuum map, not shown but obtained and analysed as a part of the products of the FP datacube) and (ii) a varying part produced by the H$\alpha$ line (H$\alpha$ integrated flux map). The continuum is computed by taking the mean signal outside the emission line. The H$\alpha$ integrated flux map was obtained by integrating the monochromatic profile in each pixel. Strong OH night-sky lines passing through the filter were subtracted by determining the level of emission away from the galaxies \citep{Laval87}. In order to improve the signal-to-noise (S/N) ratio we used both spectral (rectangular, three channels) and a box car (5$\times$5 pixels) spatial smoothing in order to minimise the smoothing effects of the spatial resolution. The velocity estimation is done by a barycentric measurement of the velocity from the identified emission profile. The dispersion velocity has been estimated by computing the FWHM of the profile, dividing by 2$\times$(2ln2)$^{1/2}$ and correcting from the instrumental and thermal broadening. The instrumental broadening,  estimated using the calibration lamp, is  $\sigma_{inst} =$ 18 km\,s$^{-1}$.  We used $\sigma_{th} =$ 9.1 km\,s$^{-1}$ from \citet{Bordalo2009} and the previous value of   $\sigma_{inst}$. 
\begin{table}
\begin{minipage}[t]{\columnwidth}
\caption{Fabry-Perot observational setup}
%\caption{Instrumental setup}
\label{FP}
\centering
\renewcommand{\footnoterule}{}  % to avoid a line before footnotes
\begin{tabular}{lc}
\hline \hline
Fabry-Perot Parameters & Values\\
\hline
\hline
Telescope & SPM 2.1m   \\
Date   & 13$^{th}$ March 2013 \\ 
Seeing\footnote{ Seeing estimated via an unbinned direct image having a pixel scale of 0.33"/px}& 1\farcs5 \\
Instrument  & PUMA-FP  \\
Detector    &   CCD Marconi\\
Pixel size    & 1\farcs3/pix  \\ 
Calibration lamp ($\lambda$) & 6562.78 {\AA}   \\
Resolution ($\lambda/\Delta \lambda$)&  8031 \\
Filter Central Wavelength          &  6607 \AA\\
Filter Transmission                     &  66\%@6586\AA        \\
Filter FWHM ($\Delta \lambda$)&  89\AA\\
Finesse$@$H$\alpha$    &   24\\
Interferometer order$@$H$\alpha$  & 330 \\
Free spectral range$@$H$\alpha$  & 19.7\AA --  911\,km s$^{-1}$ \\
Number of scanning steps   & 47 \\
Sampling step &  0.43\AA --19.38 km s$^{-1}$ \\
Interferometer FWHM ($\delta \lambda$) &  0.82 \AA \\
Total Exposure Time (sec)               &  11280   \\ 
\hline
\hline
\end{tabular}
%\footnote{ Seeing estimated with an unbinned direct image with a pixel scale of 0.33"/px}

\end{minipage}
\end{table}

%-----------------------------------figure2----------
%-------------------------- end figure 2 UVOT- SDSS profiles------------

\section {The NGC~3447  system:  photometric and kinematic properties}
\subsection{Photometric properties}\label{subsec:phot-prop}
The UV integrated magnitudes,  corrected by galactic extinction are given in
Table \ref{exptimeUV}. Figure \ref{SDSSprof} (left panel) highlights the UV luminosity profiles.
These profiles have been fitted using an  almost exponential disk (n=1.4). The model, given by a S\'ersic law, 
has been subtracted from the original profiles and residuals are shown in the  bottom left panel of Fig. \ref{SDSSprof}, for all the {\tt UVOT} filters.  In the fit, the bar has been masked as well as the
region influenced by the instrumental PSF. For this reason the residuals are shown starting from 20\arcsec,  where the disk clearly emerges (n=1.4).
\footnote{We remind here that the Sersic law is a generalization of  the de Vaucouleurs r$^{1/4}$ law \citep{deVauc48} where the value n=4 is typical of
Ellipticals and bulges, n=1 of exponential disks \citep{Freeman70}, and n=0.5 of a Gaussian profile.}
Figure \ref{SDSSprof} shows the brightness profiles in the SDSS bands and the residuals by performing the same fit as in the UV bands. We derive n=1.4 also in this case.
 The residuals do not show any significant trend, both in the UV as well as in the SDSS bands,  in particular in the region covered by NGC~3447A (emphasised in Fig.\ref{SDSSprof}).\\
\indent
There are no U, B, V magnitudes for this galaxy  in the {\tt RC3} catalog.
The {\tt Hyperleda}  catalog reports: M$_B$=-17.28$\pm$1.11\,mag. The large error  includes uncertainties in the  observed magnitude  (13.90$\pm$0.58\,mag corrected for galactic extinction), internal extinction,  inclination effect, and distance modulus estimate ($\pm$0.94 mag).
The AB integrated magnitudes we derived from optical SDSS images (Table \ref{SDSS})  are  0.3 mag brighter, on average, than the total magnitudes of \citet[their Table 4]{Marino2010} which, as outlined by the same authors,  are estimated  in a reduced area because the system is near the edge of their SDSS image.\\
%Figure \ref{SDSSprof} shows the brightness profiles and the residuals by performing the same fit as in the UV bands. We derive n=1.4 also in this case.
\indent
The distance of NGC~3447 is very uncertain (30\%).  
 From Extragalactic Distance Database (EDD) we derive the
 largest value, 20.9\,Mpc,  whereas
following our work on LGG 225 \citep{Marino2010, Mazzei2014}, we assign to this system  a distance of $\simeq$15\,Mpc.  This choice is consistent with the value of  H$_0$=75\,km\,s$^{-1}$\,Mpc$^{-1}$  and the systemic velocity in Section 3, as well as with the other cosmological parameters, $\Omega_{\Lambda}$=0.73, and $\Omega_{bar}$=0.27 given by the NED catalogue.
This implies an uncertainty in its total absolute magnitude of  -0.72\,mag. 
Therefore, accounting for this distance, the g magnitude in Table \ref{SDSS}, and the \citet{Windhorst91} filter transformation, B=g+0.51+0.6$\times$(g-r)  giving B=13.25\,mag,
 we derive the  absolute magnitude in the B-band: M$_B$= -17.63 +/-0.72 mag. %B=12.57+0.163=12.73 +/- 0.05. m-M=30.88 (con d=15Mpc)+/-1.5 (con D=20.9 Mpc) MB=-18.15 +/-0.72 mag
Unfortunately, a recent study of the  SNa 2012ht in NGC~3447 by \citet{Yamanaka14}, which classified this SNa as a
transitional SN Ia between normal and subluminous  SNae Ia,  does not provide any distance (modulus) estimate.

 %---------------------------- figure3 Maps 
\begin{figure*}
\center
\includegraphics[width=18cm]{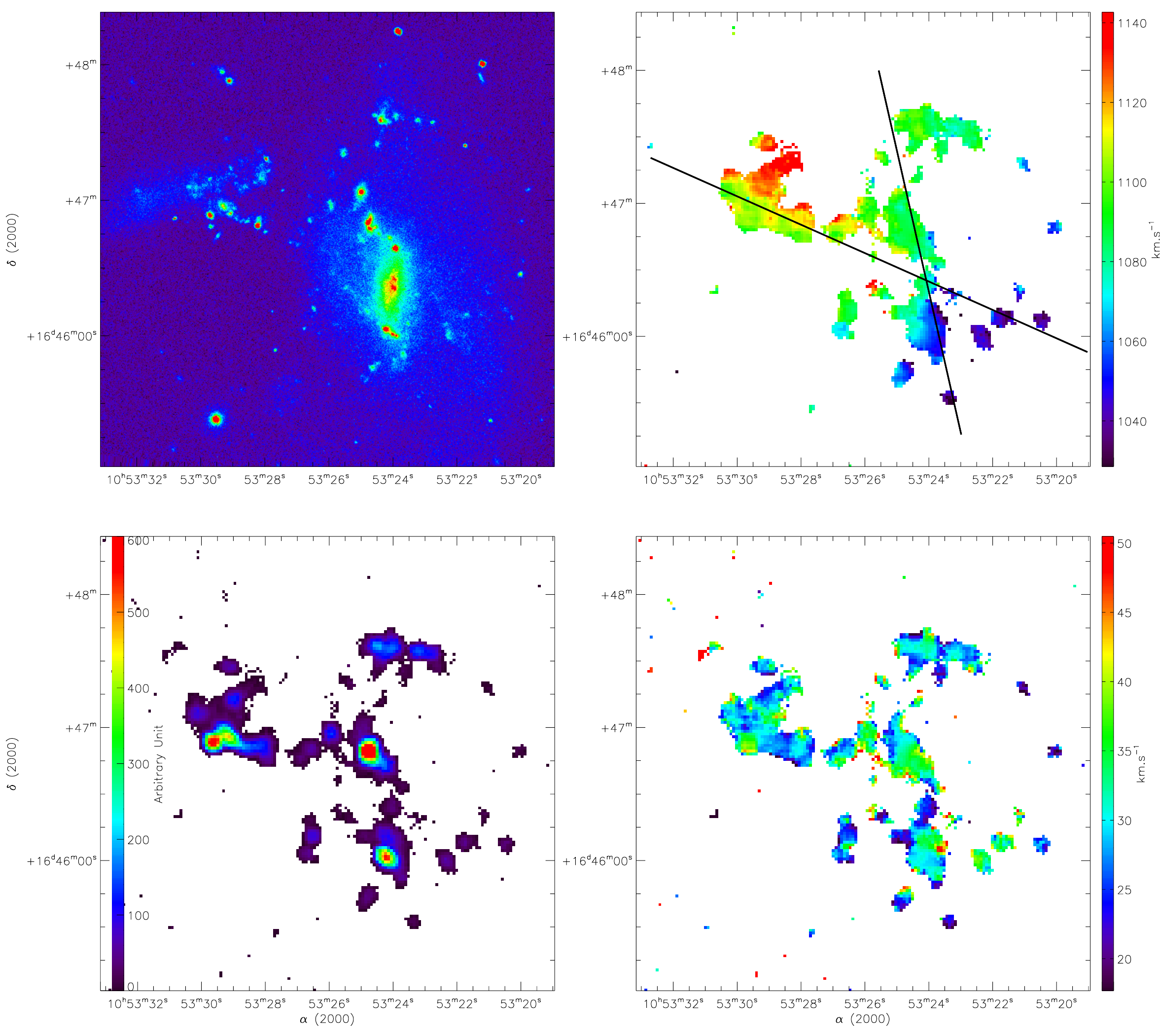}
\caption{ {\it Top:} SDSS $r$-band image of NGC~3447/3447A (left) and the velocity field of  H$\alpha$ emission (right).
{Bottom:}  the H$\alpha$, monochromatic emission map (left) 
and its velocity dispersion map (right). All panels correspond to the same  3\farcm5$\times$3\farcm5  field,  and are on
 the same spatial scale with North on the top and  East to the left. Black solid lines in the top right panel show the position angles along which the velocity profiles, shown in Fig. \ref{fig3vel}, are derived.} 
\label{figmap}
\end{figure*}
%-------------------------- end figure3 Maps 

%---------------------------- figure 4 Vel  profiles
%\begin{minipage}[t]{\columnwidth}
\begin{figure*}
\center
\includegraphics[width=16cm]{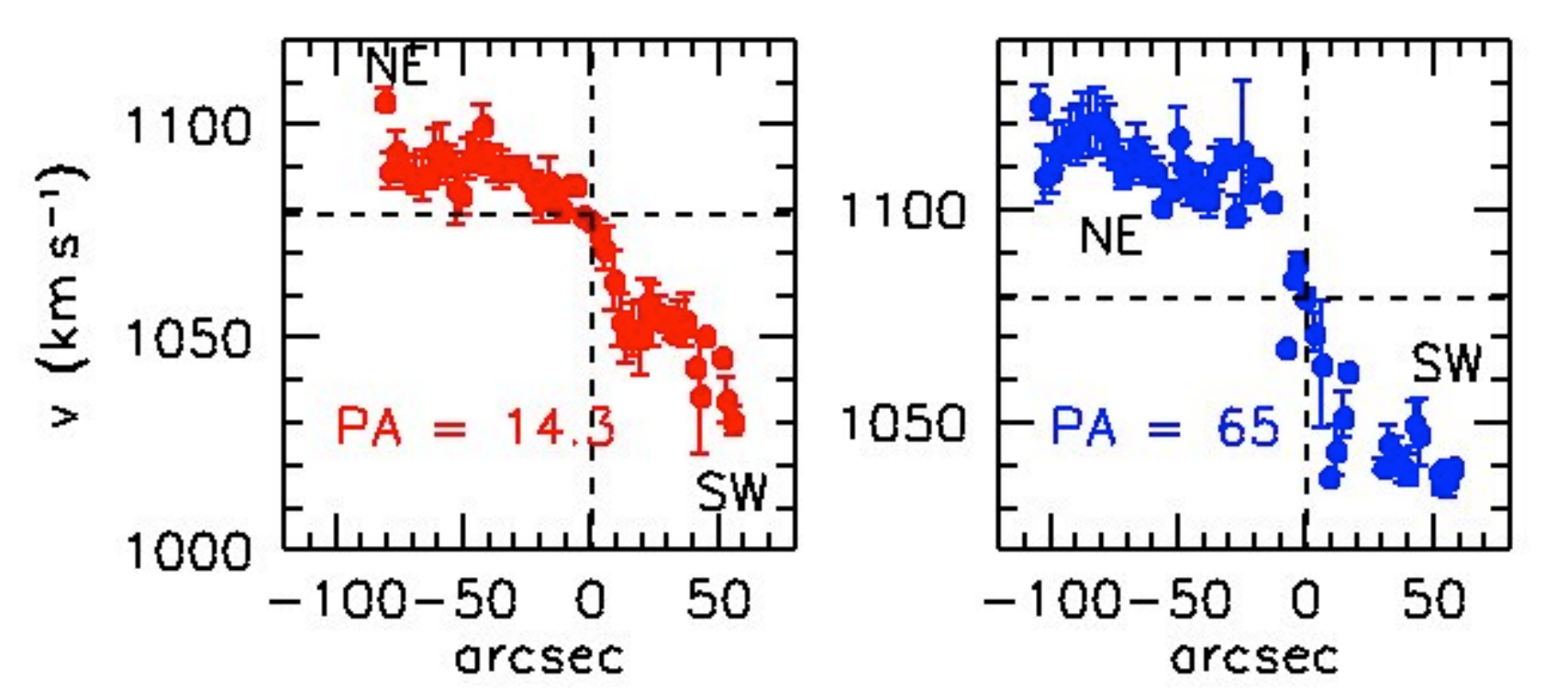}
\caption {Velocity profiles along the lines of sigth in Fig.\ref{figmap} (top left panel) 
corresponding to  PA=14.3$^o$, and PA=65$^o$,  which connects  NGC 3447A; 
horizontal dashed lines indicate the derived systemic velocity (1079\,km\,s$^{-1}$, Section \ref{subsec:Fabry-Perot observations}).  
}
\label{fig3vel}
\end{figure*}
%\end{minipage}
%-------------------------- end figure 4 Vel profiles
\subsection{2D H$\alpha$ kinematic analysis: moment maps}
\label{subsec:momentmaps}
In the top left panel of Fig. \ref{figmap}, we show the SDSS $r$-band image of the same field observed with
FP.  
%\label{fig2map}a shows first an SDSS image in the r band, \label{figXX}b shows the velocity map, \label{figXX}c shows the H$\alpha$ monochromatic map and \label{figXX}d the dispersion velocity map.
The velocity map of NGC~3447 (top right panel) does not show any regular disk rotation pattern, rather a small velocity gradient, between 1043 km\,s$^{-1}$ in the southern tip and 1085 km\,s$^{-1}$ in the North. 
NGC~3447A also shows a velocity gradient between 1090 km\,s$^{-1}$ in the South East and 1130 km\,s$^{-1}$ in the North West. \\
\indent
The SDSS $r$-band centre of NGC~3447  does not correspond to any emitting region in the H$\alpha$ monochromatic map (Fig. \ref{figmap} left,  up and bottom panels). 
The same map shows three very bright emitting regions, one South and one North of the optical central bar, and the third to the South East region of NGC~3447A.\\
\indent

The velocity dispersion map (bottom right panel) shows that almost all the emitting regions have supersonic velocity dispersion, between 20 km~s$^{-1}$ and 45 km~s$^{-1}$. 
We estimated the mean velocity dispersion of the all objects is  30 $\pm$ 6 km~s$^{-1}$, outlining the highly supersonic nature of the ionised gas in this galaxy.
By comparing the monochromatic emission and velocity dispersion maps, we note  a different behaviour for each of  the three brightest regions of the monochromatic map. The one in the South shows the lower velocity dispersion with respect to its surrounding, the  second, in the East, shows the opposite trend, and the third, in the centre, does not show any clear pattern.  The behavior of the southern region is the same found in several studies of Giant  H\,{\textsc{ii}} Regions and  H\,{\textsc{ii}} galaxies \citep{Munoz-Tunon1996, Bordalo2009} and in interacting pairs \citep{Plana17}. 
  No definite answer has been firmly established  to this feature, may be due  to  the presence of expanding shells \citep{TT93}.
We performed a check of  several profiles in order to find out if multiple components could be present. We found that the profiles are fairly symmetric 
with no obvious multiple components.  This could come from the superposition in the line of sight of different HII regions/clouds.
Due to our insufficient spectral (due to the spectral smoothing in order to gain S/N) and  spatial resolution, no further conclusions can be drawn.\\
\indent
With the purpose to constrain our simulation,  we obtained the radial velocity position profiles, shown in Fig. \ref{fig3vel} ,  along the lines of sight highlighted in Fig. \ref{figmap} (top right panel), corresponding to  (i)  the  North Eastwards  major axis position angle (PA),  PA = 14.3$^o$ (Hyperleda catalogue), and  (ii)  PA=65$^o$,  the line of sight connecting the centre of NGC~3447, selected from the r-band SDSS image, to NGC~3447A, 
both accounting for a semi-angle sector of 25$^o$ \citep[their Figure 9c]{Amram2007}. The velocities are estimated in different crowns of two pixels each.\\
\indent
 To summarise, we do not find any kinematic signature probing that NGC~3447 and NGC~3447A have different velocity fields. 
The  velocity field does not show any rotation pattern.  The velocity gradient across the entire system is too weak and irregular to be considered as a rotation pattern,  although the South West region appears slightly approaching and the North East slightly receding.  
Our results point towards the idea that  patchy and distorted features are different regions of a single system.

 %------------ table 4 input SPH---------------------
\begin{table*}
%\scriptsize
\center
\caption{Input parameters  of SPH-CPI simulation of NGC 3447/NGC~3447A}
\begin{tabular}{llccccccccc}
\hline
 N$_{part}$&a &p/a& r$_1$ & r$_2$ & v$_1$ & v$_2$ & M$_T$ & f$_{gas}$\\
   &[kpc]&  & [kpc]&   [kpc] & [km/s] & [km/s]  & [$10^{10}\,M_\odot$] & \\
\hline\hline
 80000 &376 & 1/3 & 505  & 505 &18&18& 20& 0.1   \\
\hline
\end{tabular}

\footnotesize{Columns are as follows: (1) total number of initial (t=0) particles, N$_{gas}$=N$_{DM}$=N$_{part}$/2;
(2) length of  the semi-major axis of the halo; (3) peri-centric separation of the halos in units of the semi-major axis; (4)  and (5)  distances of the halo centres of mass from the centre of mass of the total system, (6) and (7) velocity moduli of the halo centres in the same frame; (8) total mass of the simulation; (9) initial gas fraction of each halo.}
\label{tablesim}
\end{table*}
%-------end table 4 input SPH--------------------- 
\section{Comparison with simulations}
\label{sec:simul}
We  explored the evolutionary scenario of the NGC~3447/NGC~3447A system  exploiting a large
set of SPH-CPI simulations  of galaxy encounters and/or mergers, including a chemo-photometric code based on evolutionary population synthesis (EPS) models.
The general prescriptions of our SPH-CPI simulations are reported in several previous papers \citep[and references therein]{Mazzei2014, Mazzei2014_2, Buson2015}.\\
\indent
All the simulations start from collapsing triaxial
systems composed of dark matter (DM) and gas in different proportion and different total masses, as in \citet[][MC03 hereafter]{MC03}.
 The simulated  halos have all the same initial conditions, i.e.,  virial ratio (0.1),  
average density, and spin parameter.
 In more detail, each system is built up with a spin parameter, $\lambda$, given by $|{\bf J}||$E$^{0.5}$/(GM$^{0.5}$), where E
is the total energy, J is the total angular momentum, and G
is the gravitational constant; $\lambda$ is equal to 0.06 and aligned with
the shorter principal axis of the DM halo. The initial triaxiality ratio of
the DM halos, $\tau$ = (a$^2$ - b$^2$)/(a$^2$ - c$^2$),  is 0.84  \citep{Mazzei2014, Mazzei2014_2} where a $>$ b$ >$ c
\citep{Warren1992}.
 This $\tau$ value is different from the fiducial value of  $\tau$=0.58
adopted by MC03  and
motivated to be closer to the initial  condition  of cosmological halos   as in \citet{Warren1992}, 
\citet[][their Table 1]{CMM06}, and \citet{Schneider12}.\\
\indent
The simulations include self-gravity of gas, stars and DM, radiative cooling, hydrodynamical pressure, 
shock heating, viscosity, star formation, feedback from evolving stars and type II supernovae, and chemical enrichment.
%---------------------------- figure 5 ---------------------------------------------
\begin{figure*}
\centering
\includegraphics[width=15.cm]{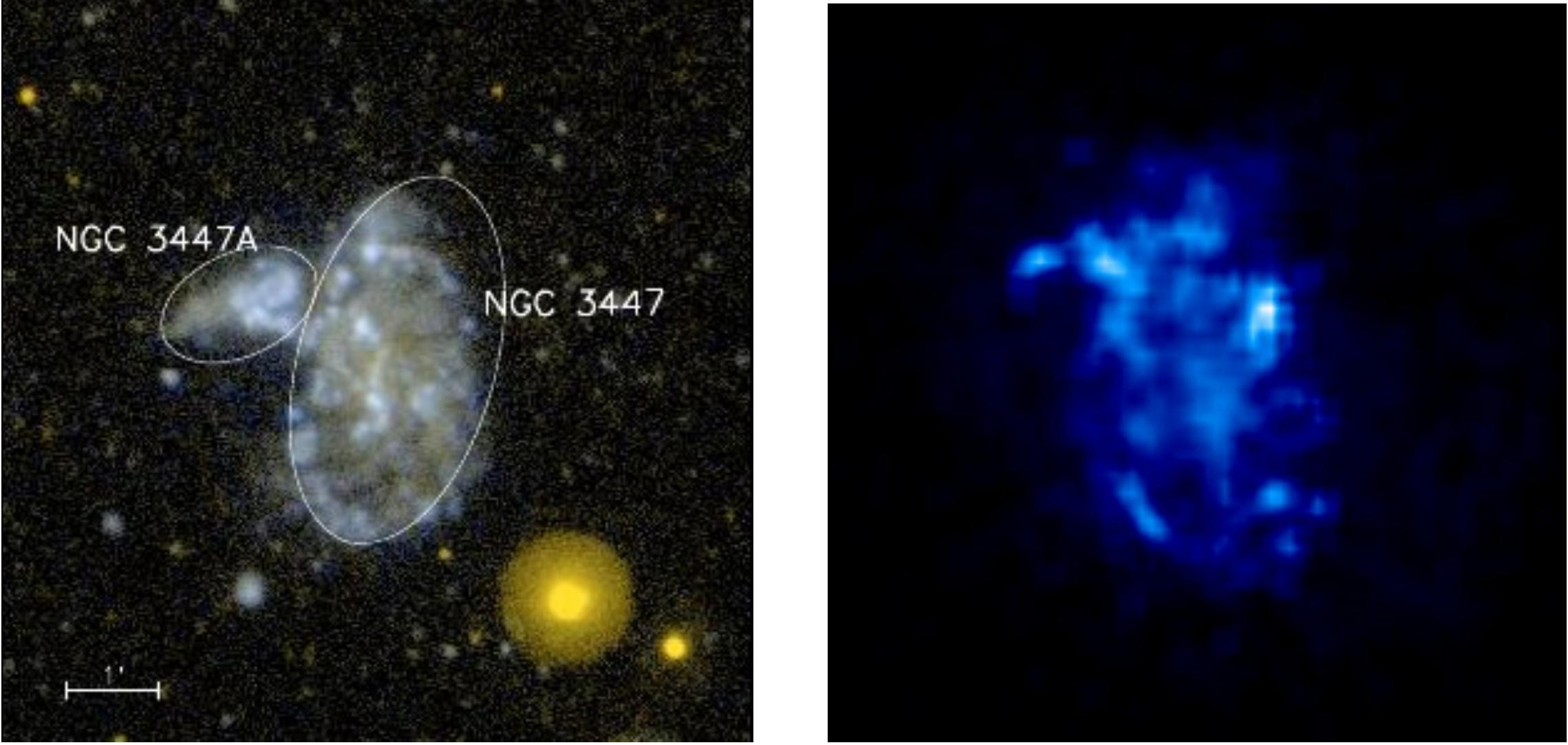}
\caption{Comparison between the GALEX NUV (yellow), FUV (blue) 
composite image (left panel), 7\arcmin$\times$7\arcmin, from \citet{Marino2010} and the FUV band luminosity density map  (XY projection) from the snapshot (right panel)  best fitting the global properties of NGC~3447 normalized to the total flux in the box; maps are on the same spatial scale, and resolution (5\arcsec). } 
\label{fig4}
\end{figure*}
%-------------------------- end figure5  ---------------------------------------
%---------------------------- figure 5-bis ---------------------------------------------
\begin{figure*}
\centering
\includegraphics[width=4.5cm]{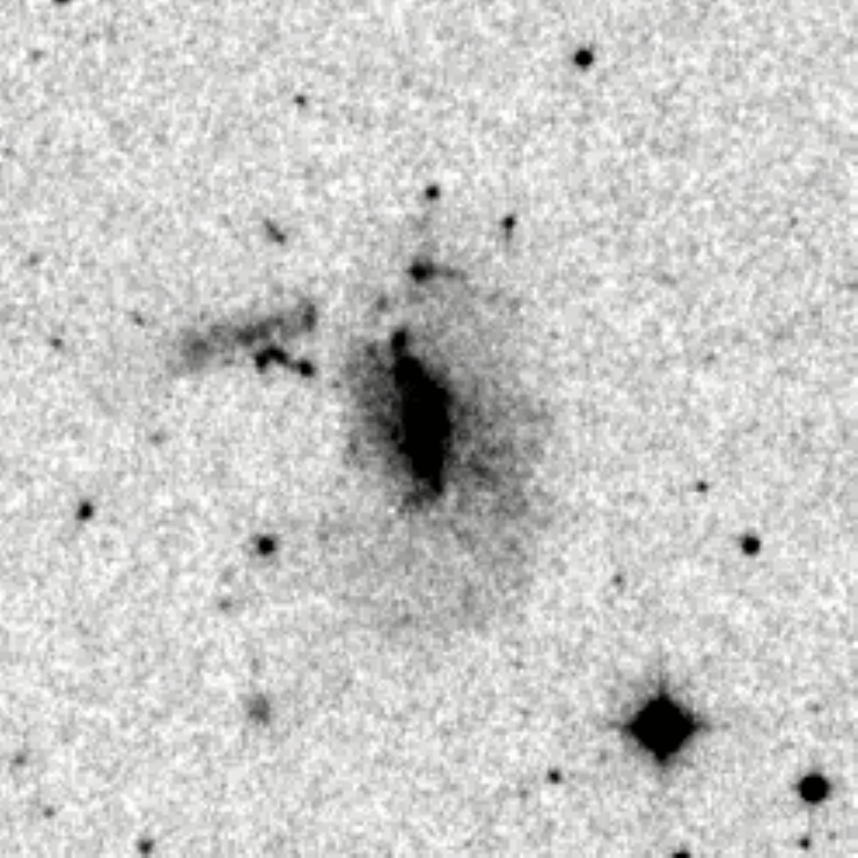}
\includegraphics[width=4.5cm]{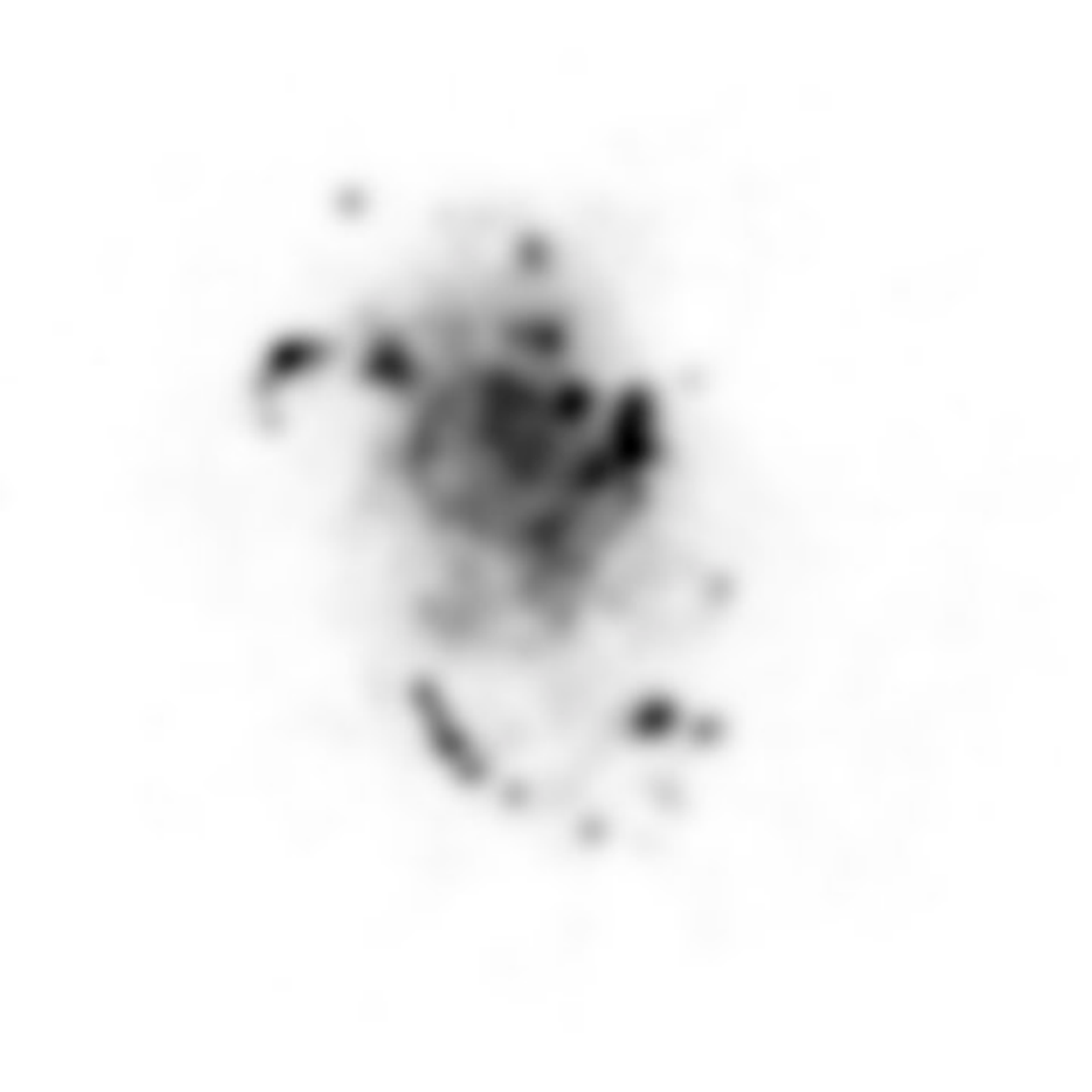}
\includegraphics[width=4.5cm]{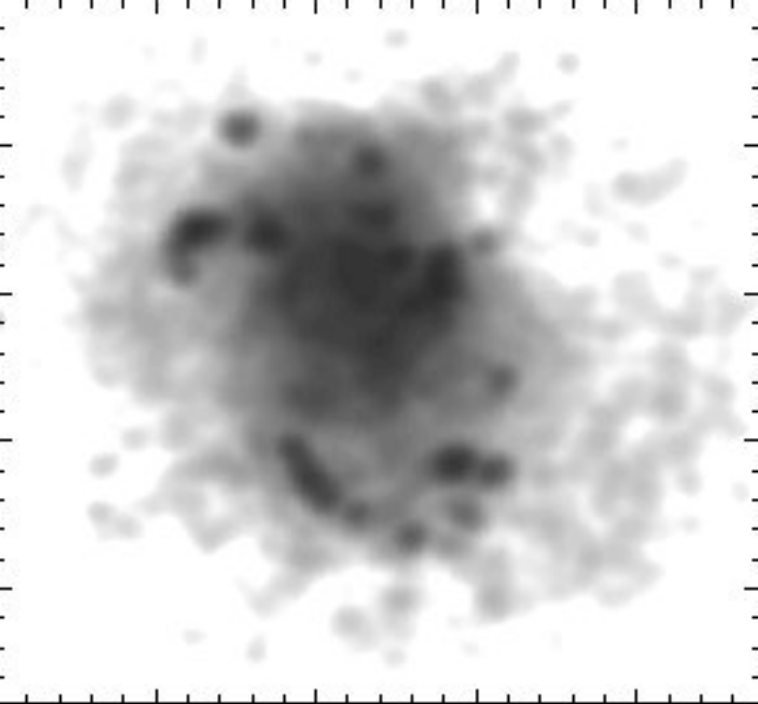}
\caption{Comparison between the B-band DSS image from NED (left panel), 7\arcmin$\times$7\arcmin,  and the B-band luminosity density maps  (XY projection) from the snapshot ( middle and right panels)  best fitting the global properties of NGC~3447A/NGC~3447 normalized to the total flux in the box; maps are on the same spatial scale, and resolution, (1\farcs7)   and with  different contrast, 200 (middle) and 1500 (right).
}
\label{fig4b}
\end{figure*}
%-------------------------- end figure5-bis  ---------------------------------------
%---------------------------- figure 6 ---------------------------------------------
\begin{figure*}
\center
\includegraphics[width=4.5cm]{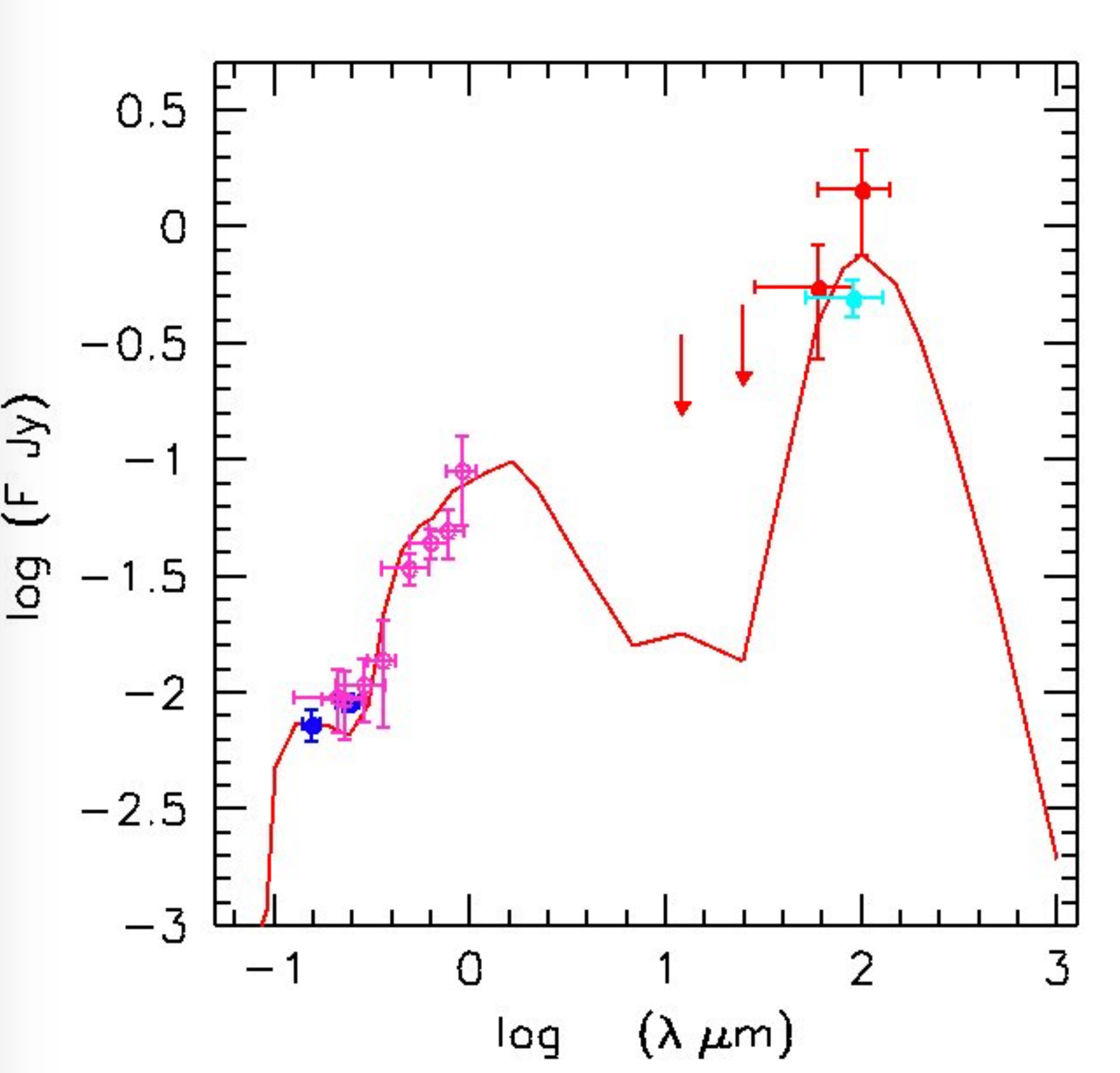}
\includegraphics[width=4.5cm]{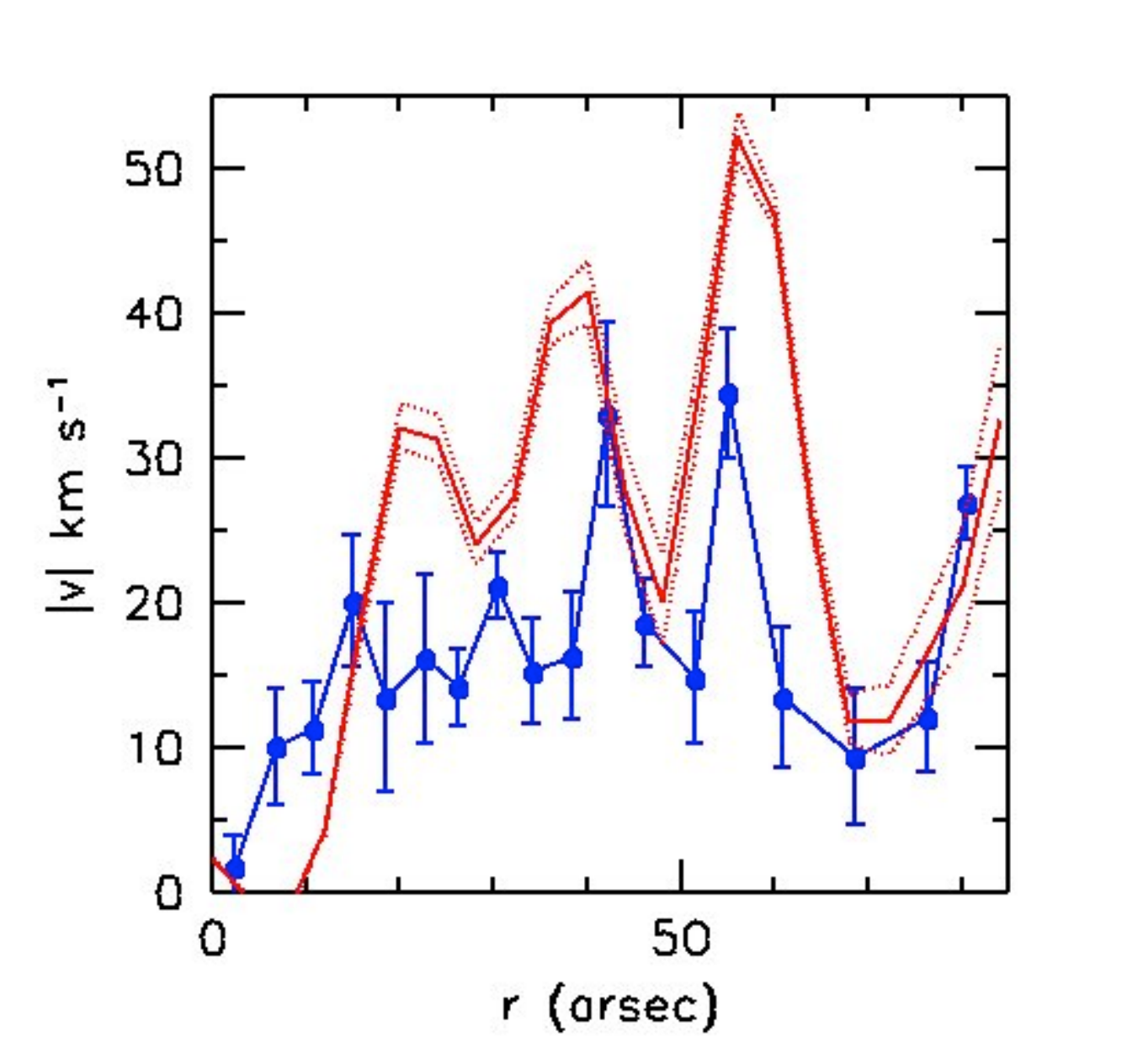}
\includegraphics[width=4.5cm]{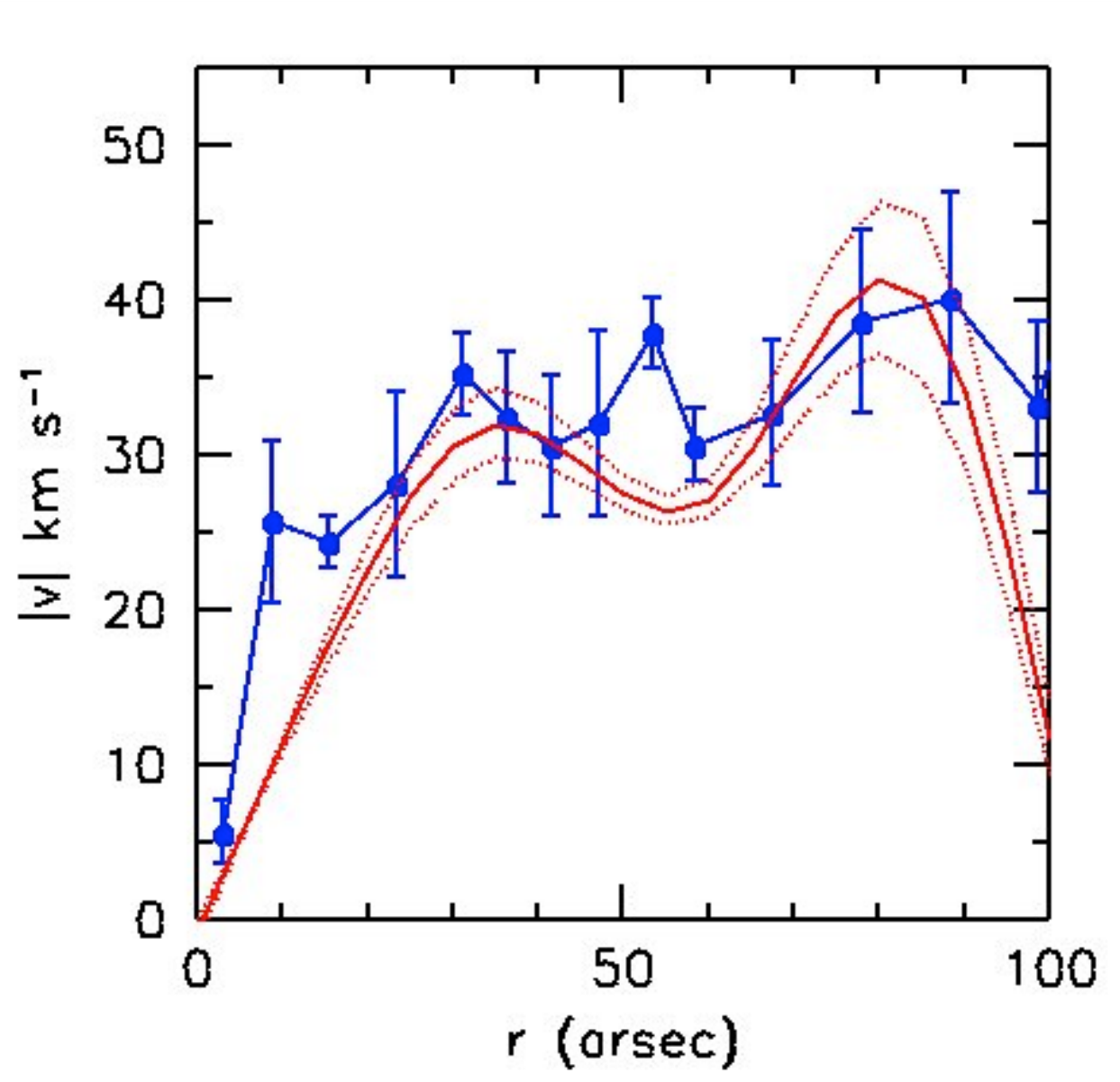}
\caption{
{\it Left:} The measured (filled dots)  and the  predicted (red  solid line) SED of the  NGC~3447A/NGC~3447 system;
blue, magenta,  red, and azure dots are respectively  for far-UV {\tt GALEX}, from  \citet{Marino2010},  {\it Swift}-{\tt UVOT} and SDSS  (Table \ref{exptimeUV} and \ref{SDSS}),  IRAS and {\tt AKARI-F} \citep[Catalog Archive Server, ][]{Yamauchi2011}  data respectively.
{\it Middle and Right:}  Comparison between the observed velocity profiles along the directions in Fig.\ref{fig3vel}, PA=14.3$^o$ (middle) and PA=65$^o$ (right),
 with those derived from the selected snapshot in the corresponding directions and within the same semi-angle sector (i.e. 25$^o$). The velocity profiles are folded and rebinned within 5\arcsec (blue filled dots) to be compared  with the same slices of the  simulation  (red solid  and dotted lines), with the same binning.
} 
\label{fig3}
\end{figure*}
%-------------------------- end figure 6--------------------------------------------
%---------------------------- figure 6 ---------------------------------------------
\begin{figure*}
\center
\includegraphics[width=6.5cm]{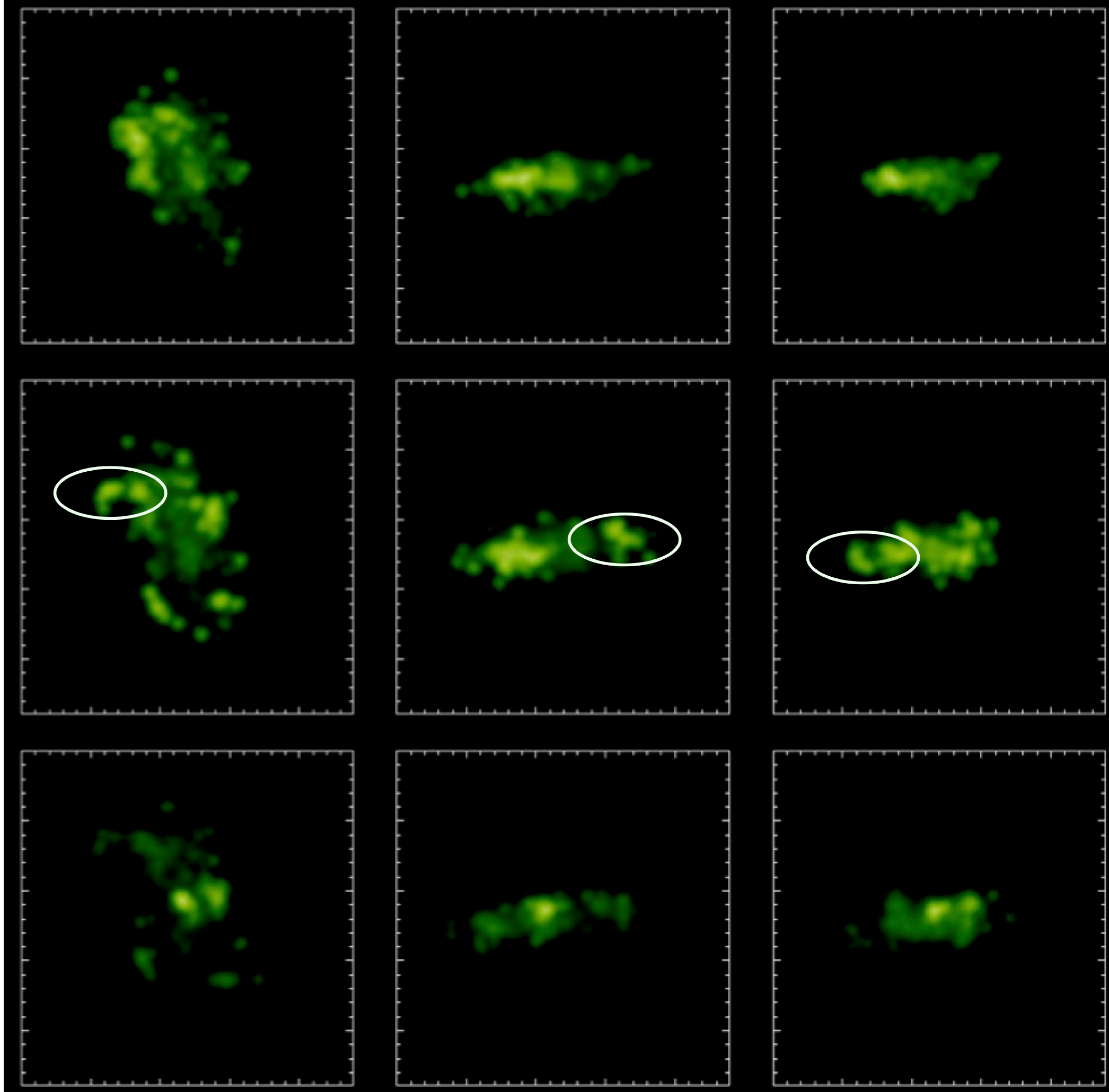}
\includegraphics[width=6.6cm]{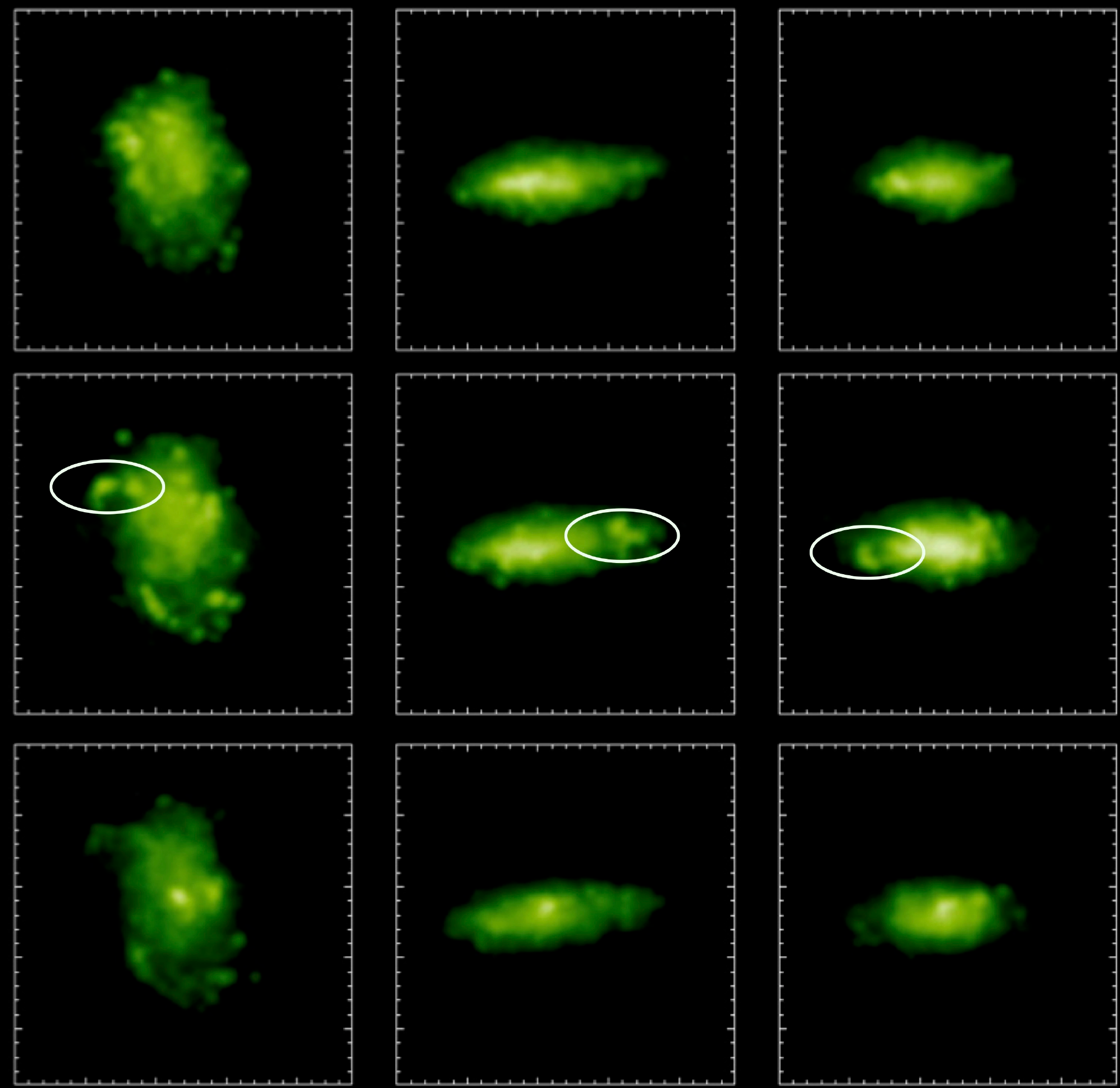}
\caption{
Projected luminosity density maps, xy, yz, and xz from left to right,  with the same filed of view and   resolution as in Fig. \ref{fig4b}. The best-fit snapshot (middle) highlights  the region corresponding to NGC~3447A (white ellipse),  the snapshot  37$\times$10$^6$\, yr after is at top,  and  that 37$\times$10$^6$\, yr  before  down.
{\it Left:}   the projected maps in the M2 {\tt Swift} filter and {\it right}, in the V-band, each normalised to the total flux inside each box.
} 
\label{newmorf1}
\end{figure*}
%-------------------------- end figure 6--------------------------------------------
%---------------------------- figure 6 ---------------------------------------------
\begin{figure*}
\center
\includegraphics[width=6.7cm]{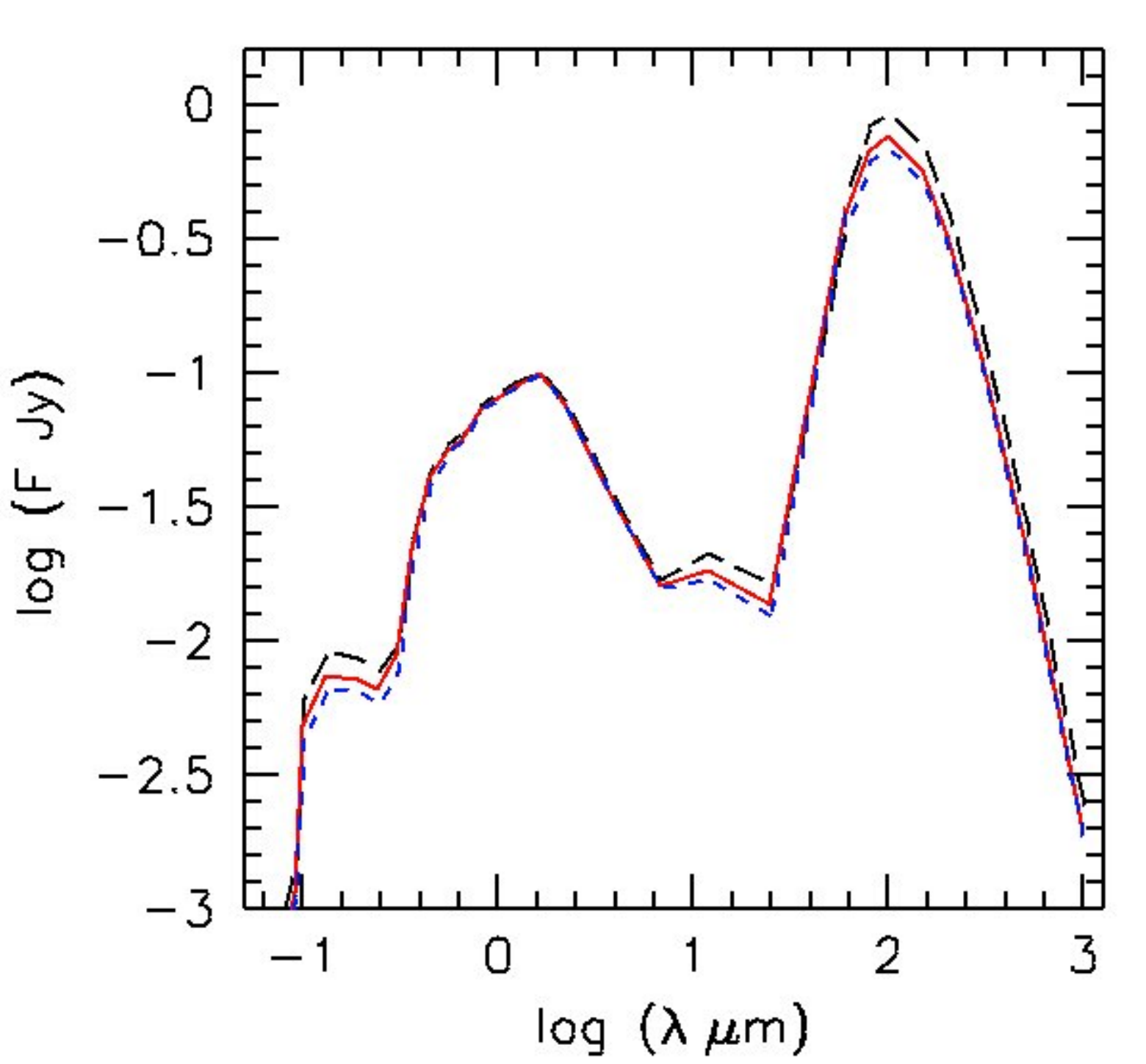}
\includegraphics[width=6.6cm]{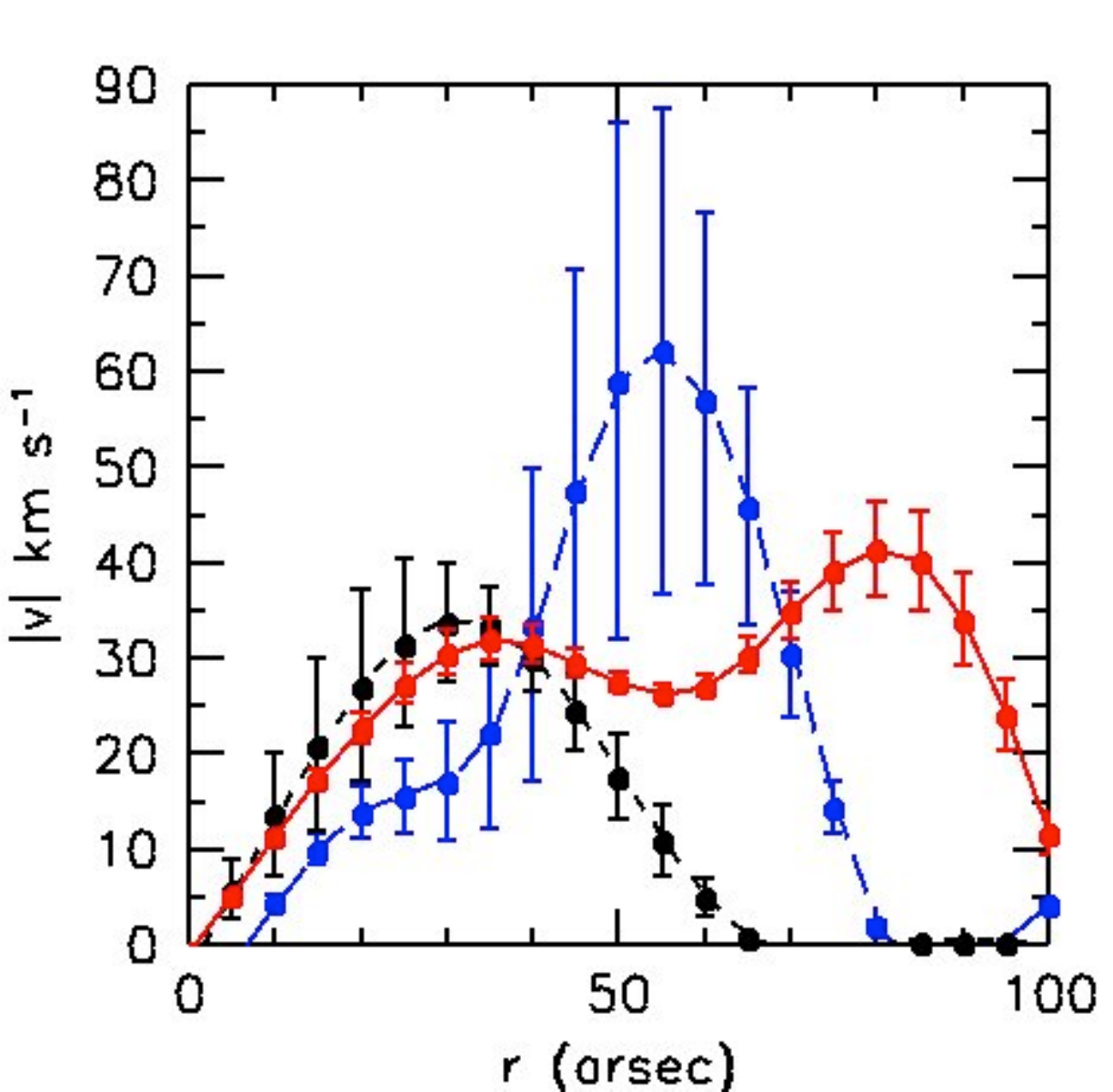}
\caption{
Comparison of the SEDs (left), and the velocity profiles (right) along PA=65$^o$, of  the best-fit snapshot (red) solid line, with those of one  snapshot  (0.037\,Gyr)  before, (blue) short-dashed line, and one beyond,
(black) long-dashed lines.
} 
\label{newsedvel}
\end{figure*}
%-------------------------- end figure 6--------------------------------------------

%---------------------------- figure7 
\begin{figure*}
\center
\includegraphics[width=6.9cm]{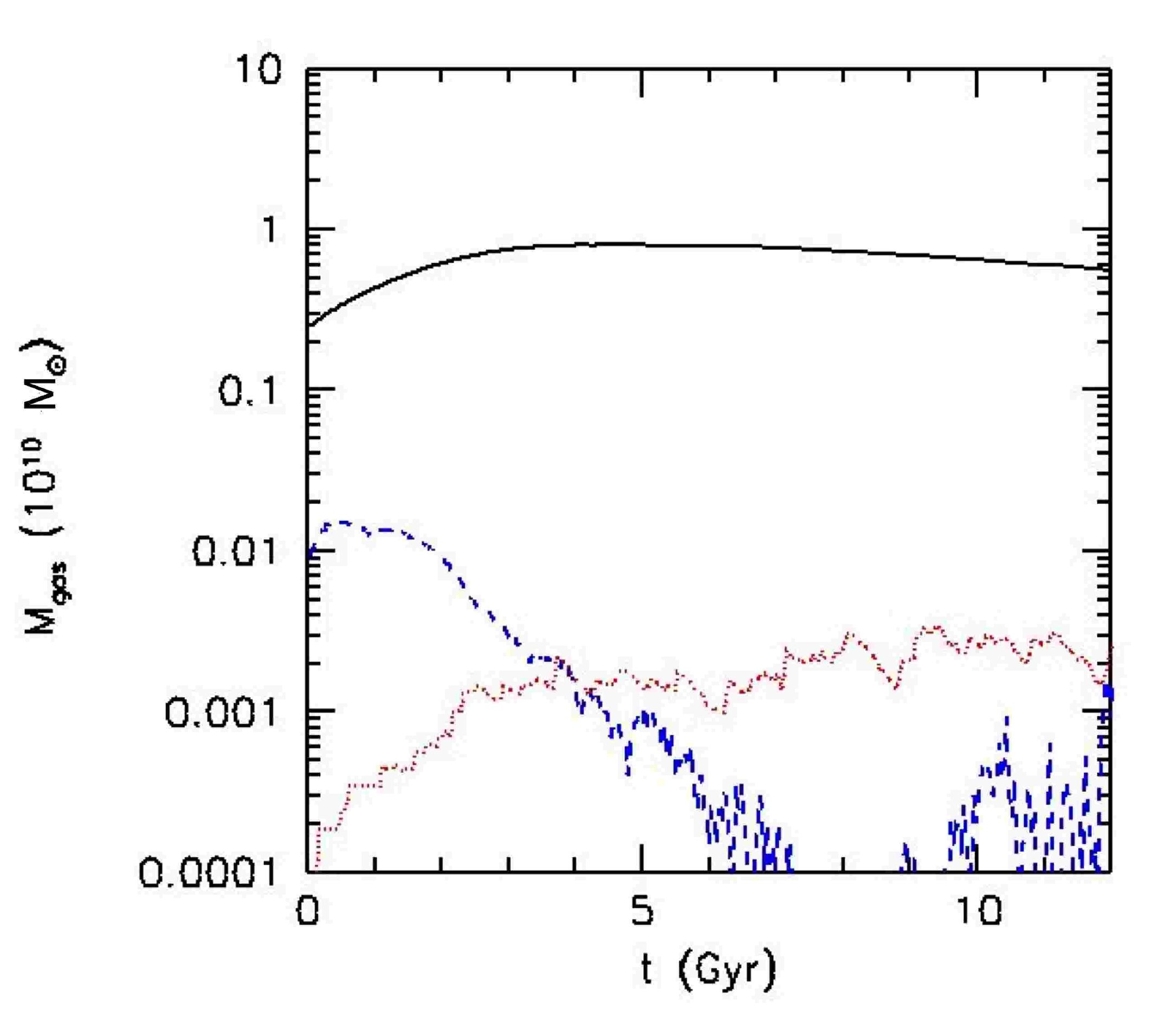}
\includegraphics[width=6.2cm]{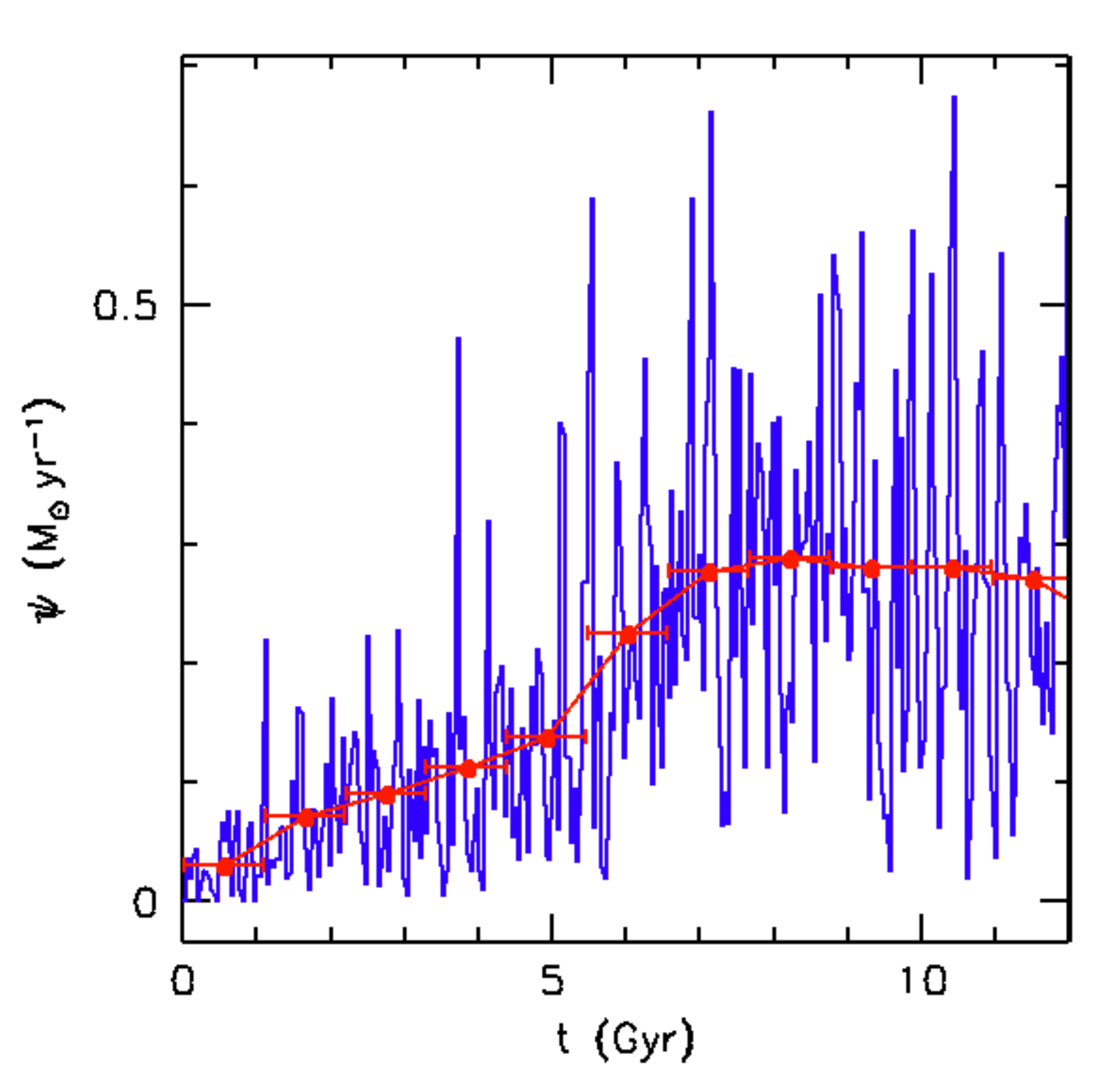}
\caption{
 The evolution of different system properties computed inside a spherical radius of 50\,kpc on the B-band luminous centre of the simulation matching the global properties of   the NGC~3447A/NGC~3447 system. Both panels stop at the actual age of the galaxy, i.e. 12\,Gyr (Section 4).
{\it Left: } The gas accretion history;  the (blue) dashed  and (red) dotted lines correspond to the gas with temperature $\le$10$^4$\,K  and $>$10$^6$\,K respectively. {\it Right:} The star formation rate (blue solid line) and its behaviour averaging within 1\,Gyr (red bins).
} 
\label{fig5}
\end{figure*}
%-------------------------- end figure7  -----------------------------------
The initial mass function (IMF) is of Salpeter type \citep{ Salpeter55} with 
upper and lower mass limit  100 and 0.01\,M$_{\odot}$ respectively, as in \citet{CM98} and MC03.\\
\indent
 All the model parameters had been tuned in previous papers devoted to analyse the evolution of isolated collapsing triaxial
halos, initially composed of DM and gas  \citep[and MC03]{CM98, Mazzei03}. In those papers, the role of the initial spin of the halos, their total mass and
gas fraction,  triaxiality ratio, as well as different IMFs, particle resolutions, SF efficiencies, and values of the feedback parameter, were all examined. The integrated properties of simulated galaxies, stopped at 15 Gyr, that is, their colors, absolute magnitudes, metallicities, and mass to luminosity ratios, had been successfully compared with those of local galaxies  (\citet[][e.g. their
Figure 17]{CM98}, \citet{Mazzei03}, \citet[their Figure 8]{Mazzei04}). In particular, a slightly higher star formation rate (SFR) compared
with the other possibilities examined, arises from our IMF choice (see MC03 their Fig.1); this allows for the lowest
feedback strength (63\% less than that in the same simulation with the lower mass limit 0.1\,M$_{\odot}$), and for the expected rotational support when
disk galaxies are formed (MC03).\\
% As pointed out by Kroupa (2012), this slope is almost the same as the universal mass function that links the IMF of galaxies
%and stars to those of brown dwarfs, planets, and small bodies
%(meteoroids, asteroids; Binggeli \& Hascher 2007).
\indent
Each simulation self-consistently provides morphological, dynamic and photometric evolution, i.e.
the spectral energy distribution (SED) at each evolutionary time,  i.e., at each snapshot. 
The time step between individual snapshots is 0.037\,Gyr . 
The SED we derive accounts for chemical evolution, internal extinction and re-emission by dust in a  self-consistent way.  This extends at least over four orders of magnitude in wavelength, that is, from 0.06 to  1000\,$\mu$m.\\
\indent
 We point out that the resulting SFR,   the  driver of the evolution of the global properties of the simulated galaxies,  converges when the initial particle number is about 10$^4$  \citep[Figure 1 of][]{MC03,Chri2010,Chri2012}.\\
 \indent
 We performed a  grid of  simulations of  mergers and encounters  starting
 from systems built up with the same initial conditions and using the same parameters tuned in the above cited papers, as described in \citep{Mazzei2014, Mazzei2014_2}. 
 The set performed account for different  masses (from 10$^{13}$\,M$_{\odot}$ to 10$^{10}$\,M$_{\odot}$ for each system), mass ratios (from $1:1$ to $10:1$),  gas fraction (from 0.1 to 0.01), and particle number (initial number of gas and DM particles from 40000 to 220000).
By seeking to exploit a wide range of orbital parameters, we varied  the orbital initial
conditions in order to have, for the ideal Keplerian orbit of two points of given masses, the first peri-centre separation,
p, equal to the initial length of the major axis of the more massive triaxial halo down to 1/10 of the same
axis. For each peri-centre separation we changed the eccentricity in order to have hyperbolic orbits of different energy. The spins of the systems are generally parallel
each other and perpendicular to the orbital (XY) plane, so we studied direct encounters. Cases with misaligned spins have been also analysed in order to investigate the
effects of the system initial rotation on the results.\\
 \indent
 From our grid of SPH-CPI simulations we single out the simulation that simultaneously (i.e.,  at the same snapshot)  accounts for the following  observational constrains which correspond to  the global properties of  NGC~3447A/NGC~3447:
i)  total absolute B-band magnitude within the range allowed by observations (Section \ref{subsec:phot-prop});
ii)  integrated SED as the observed one; 
iii)  morphology like the observed  one in the same bands and with the same spatial scale; iv) velocity field like that derived (Section \ref{subsec:momentmaps}).
The results we present are predictions from the simulation which best reproduces all the previous observational constrains at the same snapshot. This snapshot sets the age of the galaxy.\\ %We point out that the onset of the star formation is delayed compared to the begin of the simulation depending on the total mass of the halo (MC03) and the properties of the merger/encounter, so the age of the simulation is older than the galaxy age.\\
\indent
We address the reader to the papers by \citet{Mazzei2014} where our approach has been applied to early-type galaxies of two groups, USGC~367 and LGG~225, and \citet{Mazzei2014_2} where our approach has been exploited to match not only 
photometric, but also structural (e.g. disk versus bulge) and kinematical (gas versus stars) properties of two S0 galaxies, namely NGC~3626 and NGC~1533, to predict their evolution (\citet{Spa09,Tri12,Bettoni12,Buson2015, Plana17}).\\
\indent
The initial conditions, i.e. particle number, total mass, gas fraction, together with orbital parameters, of the selected simulation which corresponds to an encounter, are given in Table \ref{tablesim}. 
Moreover, the initial spin of the halos are perpendicular; 
%the initial gas fraction is the same in both the systems. 
%This value is very similar to the value of 0.13 found by \citet{Gonzalez2013} by analyzing a large sample of clusters with different total masses.
the initial gas mass resolution is 5$\times$10$^5$\,M$_{\odot}$, that of DM particles is nine times larger, and that of the star forming clusters 2$\times$10$^4$\,M$_{\odot}$.
The gravitational softening is 1, 0.5, and 0.05 kpc for DM,  gas and star particles respectively. 
The final number of particles at least doubles the initial number in Table \ref{tablesim}.

\subsection{Matching the NGC~3447A/NGC~3447 system properties} 
\label{sec:subsimul}
We find that the simulation which best fits the global properties of the whole system,  NGC~3447/NGC~3447A,  i.e., its total B-band absolute magnitude, integrated SED (Fig.\ref{fig3}, left panel), morphology (Fig.\ref{fig4} and Fig.\ref{fig4b}), and dynamical properties (Fig.\ref{fig3}, middle and right panels),
corresponds to an encounter, not to a merger, between two halos of equal mass, each of 10$^{11}$\,M$_{\odot}$ and therefore to a total mass of 2$\times$10$^{11}$\,M$_{\odot}$ (Table \ref{tablesim}).
The snapshot which best-fits all the global properties of NGC~3447/ NGC~3447A gives   M$_B$=-17.25\,mag and a galaxy age of 12\,Gyr. However, the average age of the stellar populations is younger,  increasing from 3 to 7\,Gyr  as increasing the radius up to 15\,kpc. Furthermore, the 
age of stellar populations weighted by B-band luminosity is 1.3\,Gyr,  and 0.9\,Gyr  weighted by U-band luminosity,  both are almost constant  within the galaxy. 
The mass of stars in the same region as in Fig.s \ref{fig4} and \ref{fig4b}, comparing UV and optical morphologies,  is 2.2$\times$10$^9$\,M$_{\odot}$, that of the DM   13.7 times larger. 
Maps are on the same scale as observations, where 7\arcmin$\times$7\arcmin  correspond to about a  projected box (XY projection in this case)  of 40\,kpc$\times$40\,kpc  on the stellar mass centre,
accounting for a scale of 5.5 kpc/\arcmin as given by  NED  for H$_0$=75\,km\,s$^{-1}$\,Mpc$^{-1}$\citep{Marino2010} and the cosmological parameters in Section 3.1.\\
\indent
The simulation provides that the companion galaxy,  at the same time we are analysing (i.e. 12\,Gyr), lies about 300\,kpc faraway  from our target, corresponding to about one degree on the sky accounting for the previous scale. Moreover,  this is expected to be 0.5 mag fainter in the B-band.\\%, and the two galaxies are connected by a bridge of gas.\\% i centri di massa luminosi dei 2 sistemi distano r=300kpc poco meno di 1 grado con la scala corretta al VIrgo infall
\indent
%Figure \ref{fig3} (left panel) compares  the observed and predicted SEDs.
The far-IR (FIR) SED in Fig. \ref{fig3} (left panel) accounts for a diffuse dust emission spectrum composed by  warm and  cold dust, both including polycyclic aromatic molecules, as described by \citet{Mazzei1992, Mazzei1994}. Warm dust is located in regions of high density of radiation field, that is, in the neighbourhood of OB stars (H\,{\textsc{ii}} regions), whereas cold dust is heated by the interstellar diffuse radiation field. The distribution of the cold dust  emission  is the same as in the MW  \citep{Mazzei1992}, i.e., provides a disk of gas and stars, in agreement with our findings in Section \ref{subsec:phot-prop}. However,  the intensity required is 6-7 times larger than the average in our own galaxy,  and the warm/cold energy ratio is 0.1 instead of 0.3. A more detailed FIR coverage is needed to derive strong conclusions concerning these points, however the greater intensity of the diffuse radiation field we derive, agrees with the low metallicity of stars in this system compared with that in our galaxy since low metallicity stars are more luminous than stars with solar composition \citep[see for a review][]{Chiosi2007}. The average stellar metallicity provided by the best-fit snapshot is 0.0035, almost 5-6 times lower the that in the solar neighbour.\\
%The total mass inside  R$_{25}$, 1.7\arcmin following Hyperleda catalog, corresponding to  about 10\kpc, is  9.5$\times$10$^9$\,M$_{\odot}$, the B-band M/L 
%about 12\,,M$_{\odot}$/L$_{\odot}$.
\indent
Figure \ref{fig3}, middle and right panels, compares the observed  velocity profiles  along the lines of sight in Fig.\ref{fig3vel},  both folded with respect to the galaxy centre  and binned within 5\arcsec,  with our predictions in the same slices and with the same binning. We folded the velocity profiles although, as we
remark in Section \ref{subsec:momentmaps},  we did not find any rotation curve,  rather a  velocity gradient that it is matched, within errors,
 by the selected snapshot of our simulation.\\
\indent
Figure \ref{newmorf1} compares  UV and optical B-band projected morphologies  of  the best-fit snapshot with those one  snapshot   before (0.037\,Gyr)  and one  beyond,  showing that both NGC~3447A and NGC~3447  are pieces belonging to the same system.
NGC~3447A and NGC~3447 belong to  a  single, distorted and broken disk galaxy. This galaxy corresponds the the baryonic component of one halo of our simulation; the second halo, containing a galaxy 0.5 mag fainter (B-band) than this, is  about 1 degree faraway from our luminous  target. Therefore, such a galaxy cannot appear in any of the previous pictures whose field of view is 7'x7'.\\
\indent
Figure \ref{newsedvel} compares the SEDs and velocity profiles of the same snapshots shown in Figure \ref{newmorf1}, all computed with the same recipes as in Fig. \ref{fig3}.\\

\indent
To summarise, the SPH-CPI simulation we present suggests that NGC 3447 and NGC 3447A are a single system, the result  on the baryonic disk of its halo  instability enforced by the encounter with a companion
halo of equal mass. Following this view, pairs like NGC 3447 and NGC 3447A may be considered to belong to the class of the so-called "false pairs". These latter are know to lurk within galaxy associations: the pair members have similar redshifts but they are  gravitationally bound to the group, not  each other.  In this context, our system casts a glance on a possible sub-class of false pairs: although the projected shape of NGC 3447/3447A system suggests it is a pair,  according to our simulation  it results a single object. Its patchy structure is connected to the disk gravitational instability triggered by the interaction. In  loose groups, interactions are very effective and  strongly perturb the galaxy evolution. This is just the environment where  the NGC~3447A/NGC~3447 lies.
We will deepen the analysis of  the environment of this system in  Appendix \ref{sec: envir}.

\subsection{Insights into the evolution of the NGC~3447A/NGC~3447 system}
\begin{figure*}
\center
\includegraphics[width=6.2cm]{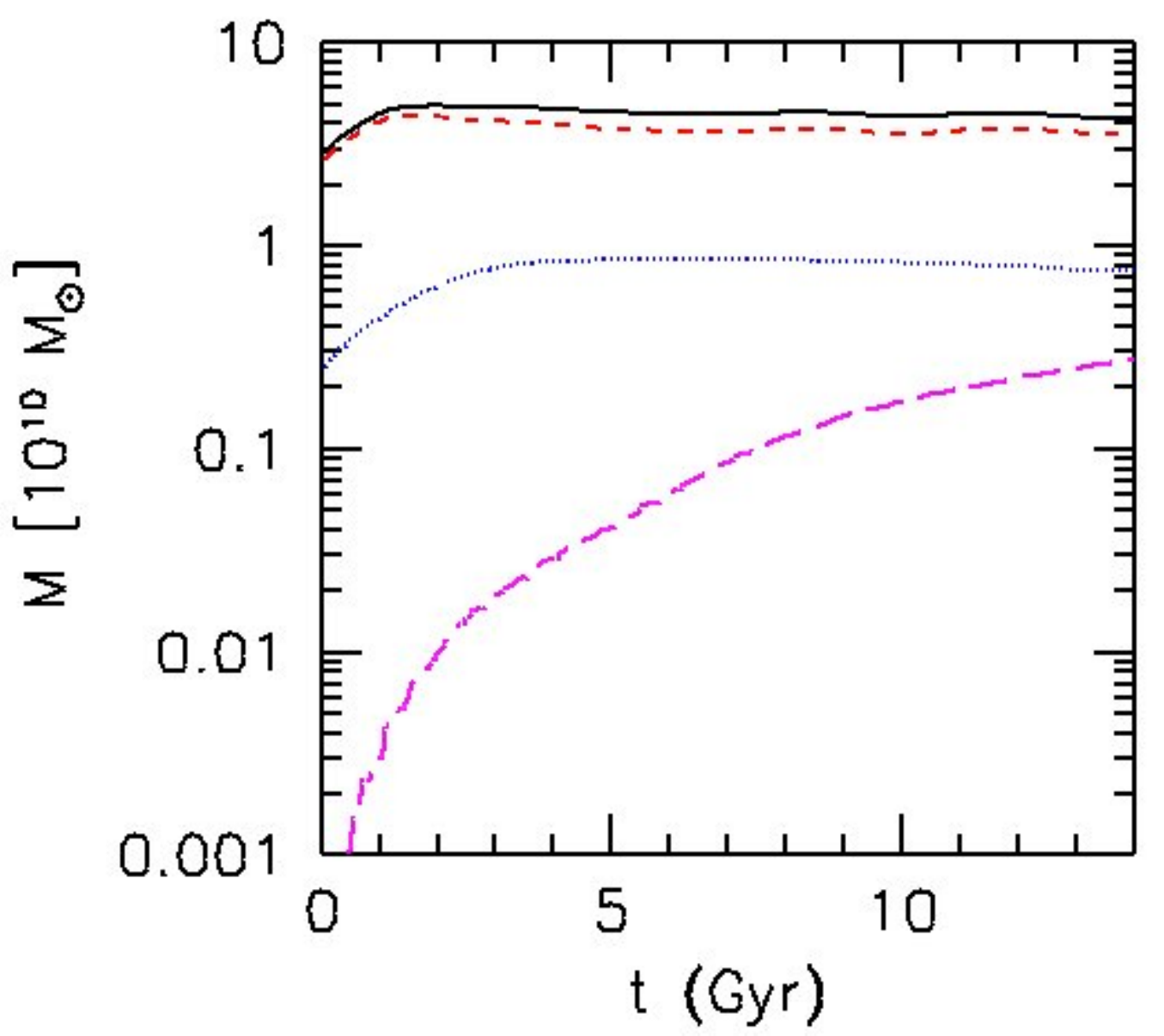}
\includegraphics[width=6.2cm]{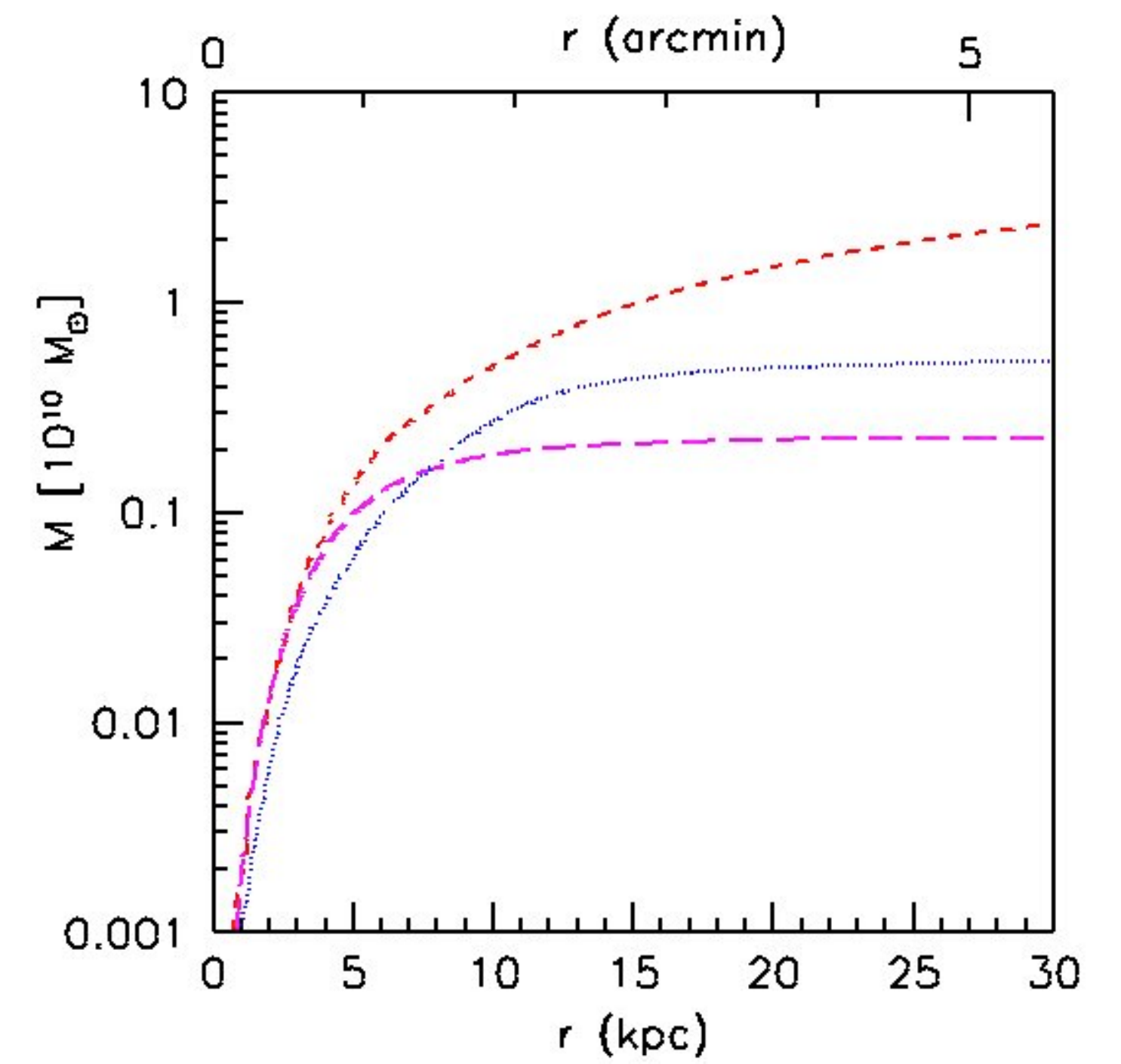}
\caption{
{\it Left:} The evolution of the different mass components within the same spherical radius of 50\,kpc  as in Fig.\ref{fig5}: (black) solid line  shows the total mass, (red) short-dashed line the DM, (blue) dotted line the gas,  and (magenta) long-dashed line the stellar mass. {\it Right:} The  mass distribution inside a spherical radius of  30\,kpc  on the B-band luminous centre, at the best-fit age: (red) dashed line  for the DM, (magenta) long-dashed for the stars, and (blue) dotted line for the gas component.
} 
\label{figmassa}
\end{figure*}

Figure \ref{fig5} (left panel) shows the gas accretion history  until the time corresponding to the predicted age of the galaxy, that is 12\,Gyr (see Section above). This figure refers to a spherical   radius of 50\,kpc on the B-band  luminous centre of the galaxy. The mass  of gas involved in the collapse within this stellar radius is lower than the total gas mass  available in the initial volume (1$\times$10$^{10}$\,M$_{\odot}$). Despite the large gas reservoir, the SFR within the same stellar radius is always below 1\,M$_{\odot}/yr$ (Fig.\ref{fig5}, right panel).\\
 The total SFR predicted by the simulation at the best-fit age,   0.28\,M$_{\odot}$yr$^{-1}$,  is in good agreement with our
estimates from the FUV luminosity following the  recipes of  \citet{Kenn98}, that is  0.24\,M$_{\odot}$yr$^{-1}$ accounting for the same IMF lower mass limit.
\citet{Pacifici2013} found that low-mass galaxies have, on average, a rising star formation history, in good agreement with our prediction. Moreover, low-mass  galaxies are susceptible to episodic (bursty) star formation events \citep{Bauer2013}, well reproduced by our simulation.\\
\indent
As discussed in several previous papers \citep[][and references therein]{MC2001, Curir08} in DM dominated disks,  as in this case (Fig. \ref{figmassa}),  the patchy and lopsided morphology  of the system, showing  a faint inner bar and an outer semi-ring, arises from the disk instability  % a lopsided instability (m=1) followed by a slowing increasing  bar instability (m=2), 
induced  by the halo gravitational instability itself, which is enhanced by the interaction. In other words, unrelaxed and/or perturbed halos allow long lasting  instabilities in DM dominated disks \citep{MC2001}. Moreover,  the barred disk configuration is made fainter by increasing the initial triaxiality ratio of the halo  (\citet{MC2001}; here 0.84, see Section \ref{sec:simul}), in agreement with the photometric properties of this system.

\section{Discussion and Conclusions}

This paper investigates the evolution of the NGC~3447A and NGC~3447 system in the
LGG 225 group. This group is a Local Group Analog, that is a group dominated by Spirals and/or a blue galaxy population \citep{Marino2010, Mazzei2014}.
The morphology of the NGC~3447A and NGC~3447 system  is  highly asymmetric, showing a ring-like structure in the outskirts, brighter in the {\it Swift} UV bands.  Its inner morphology  shows a bar,  stronger in the optical bands.  

We  obtained  multi-wavelength {\it Swift}-UV
and SDSS surface photometry, and the H$\alpha$ velocity field of the NGC~3447A and  NGC~3447 system by   performing
FP observations via PUMA@2.2-mSPM.  All these data are used to
constrain our grid of SPH-CPI simulations of mergers and encounters,  to select the simulation which best fits the global properties of both
NGC~3447A and NGC 3447.  All the other model parameters have been tuned in previous papers (Section 4).\\
We find the following results:

The fit of the surface brightness profiles from far-UV to $i$-band, using  a single S\'ersic law  starting from 20\arcsec, i.e. outside the  PSF effects and bar prevalence, suggests a  disk (n=1.4) extended up to including NGC~3447A.

The H$\alpha$ velocity field of total system does not show any regular velocity rotation pattern. We selected two suitable lines of sight to derive the radial velocity position profiles finding
no rotation along the major axis, and almost regular rotation along the line connecting NGC~3447A and NGC~3447.
%We measure a velocity gradient of about 50\, km~s$^{-1}$ and no significant velocity difference ($\Delta$V) between the galaxy center and the outer region including NGC~3447A. 

The integrated B-band absolute magnitude,  the SED extended from far-UV to far-IR, the morphology,  and the 2D kinematics of the whole system, are all best fitted by a simulation corresponding to an on-going encounter. Two halos of initial equal masses, each of 10$^{11}$\,M$_{\odot}$,  are still approaching. 
The interaction modifies the halo potentials so that the evolution of each system differs from the same in the isolated case  \citep{MC03}, and induces  the galaxy distortion we observe.

The simulation predicts: \\

(i) NGC 3447 and NGC~3447A belong  to the same halo. Moreover,  NGC~3447A is  a  substructure of the same disk that NGC~3447 belongs  to. 
 The patchy and lopsided morphology  of this system arises from the disk instability  % a lopsided instability (m=1) followed by a slowing increasing  bar instability (m=2), 
induced  by the halo gravitational instability itself, enhanced by the interaction.  Unrelaxed and/or perturbed halos allow long lasting  instabilities in DM dominated disks \citep{MC2001}, as in this case.
Therefore, the halo gravitational instability drives the disk observed morphology. The system is a "false pair" since NGC~3447A is the result of the disk instability, not a companion.\\
(ii) The age of   NGC~3447A/NGC~3447  is 12\,Gyr, that  of its stellar populations,  weighted by B-band luminosity, is  1.3\,Gyr. 
The stellar mass of this galaxy is 2.2$\times$10$^9$\,M$_{\odot}$,  its  total mass 14.7 times larger than that.\\
(iii) The  two halos involved in the interaction are separated by  about one degree now (Section 4.1). Their corresponding galaxies 
have different luminosities,  by 0.5 mag in the B-band,  the galaxy corresponding to our target, the NGC~3447A/NGC~3447 system, being the brighter one.
In Appendix A we discuss the environment of this system, where encounters are highly favoured and there are several dwarf galaxies.\\

%%---------------------------- figure 8 
%-------------------------- end figure 8  ------------------------------------
\indent
More than 25 years ago, 
in an invited review, \citet{Karachentsev1989} discussed the
incidence of {\it false pairs} in his catalog of 603 isolated pair
\citep[][KIG]{Karachentsev1972}. The incidence of {\it false pairs}, 
a class totally distinct from that
of optical pairs i.e ``accidental projection in the line of
sight of galaxies that are not in spatial proximity'',
reaches about 32\% of KIG objects. The {\it false pairs} are formed
by projection of two members of one group or cluster 
in the line of site that are not gravitationally bound to each
other. Following our simulation,  the system NGC~3347A and NGC~ 3347  
represents a new class of  {\it false pairs} that is galaxies
which appear as pairs but are a single distorted galaxy.
Karachentsev warned about the potential danger of an
underestimation of the role of {\it false pairs} since they gave
``anoumalously high values of average orbital mass for
binary galaxies''. The problem reverberates to new large
pairs surveys. Moreover, the evolution of this DM dominated system,
giving  rise to an very common intermediate luminosity galaxy as the result  of
an encounter with a still far away companion, is
the building block in  the galaxy assembly.\\

\section*{Acknowledgements}
We thank the referee for helpful comments improving our manuscript.
PM and RR acknowledge partial financial support by  contracts PRIN-INAF-2014-14 
``Star formation and evolution in galactic nuclei''. We acknowledge the use of 
public data from the {\it Swift} data archive. SDSS calibrated images of the
NGC 3347 system has been obtained from the data-base developed by \citet{Knapen2014b}
(http://vizier.cfa.harvard.edu/viz-bin/VizieR?-source=J/A+A/569/A91).
MR acknowledges financial support from grants IN103116 from PAPIIT-UNAM and 253085 from CONACyT.
HP thanks  Laboratoire d'Astrophysique de Marseille for its financial support during his visit in April-July 2017.
 We acknowledge the usage of the
 {\tt HyperLeda} database (http://
 leda.univ-lyon1.fr; \citet{Makarov2014})
 and the NED, {\tt NASA/IPAC 
Extragalactic Database} (http://ned.ipac.caltech.edu), which is operated by the Jet Propulsion Laboratory, 
 California Institute of Technology, under contract with the National Aeronautics and Space Administration

% % % % % % % % % old ref % % % % % % % % % % % % % %

\begin{appendix} %First appendix
\section{The environment of the NGC~3447A/NGC~3447A system:  updated vision of the LGC 225 group}\label{sec: envir}

\begin{figure*}
\includegraphics[width=8cm]{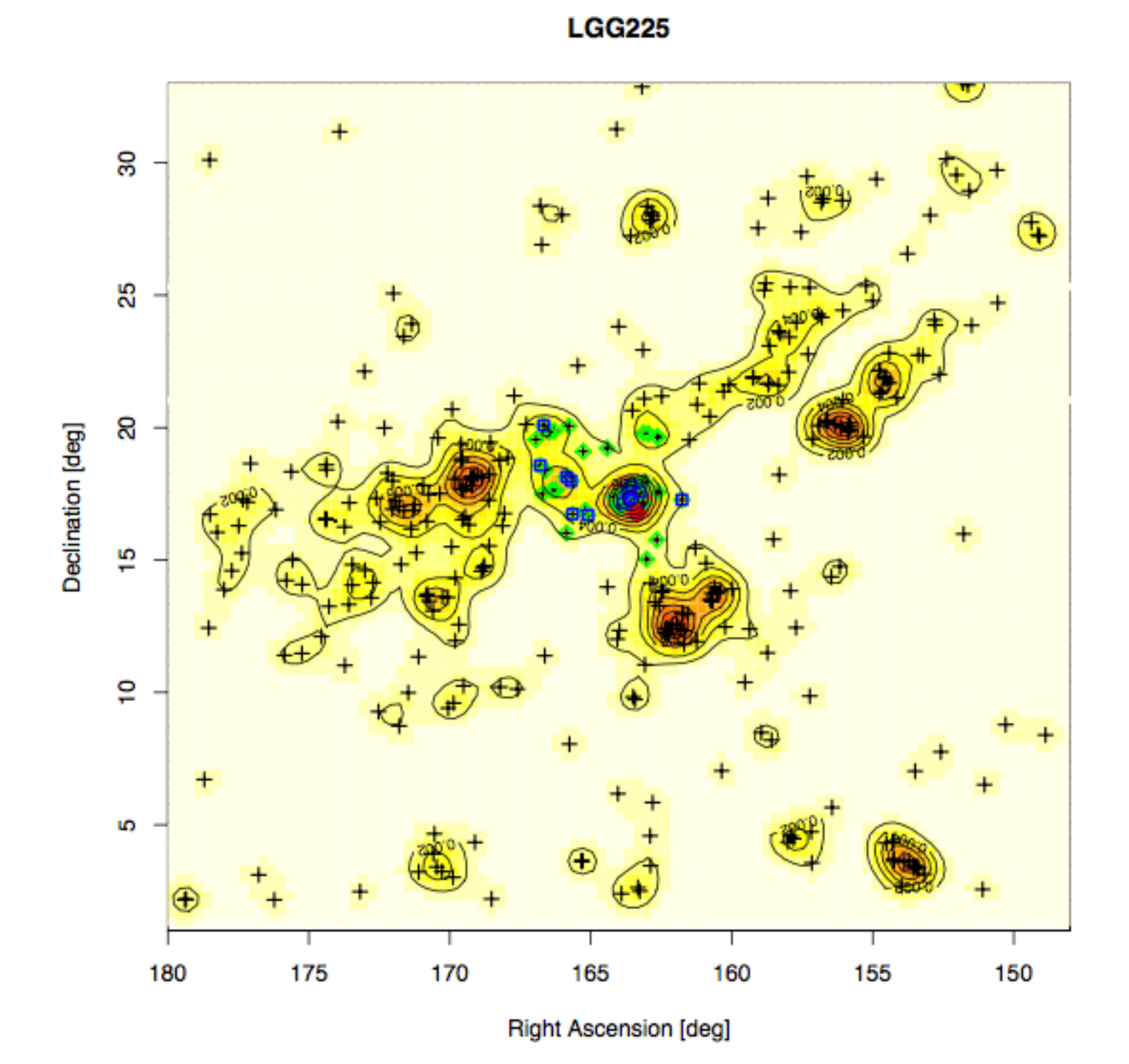}
\includegraphics[width=10.5cm]{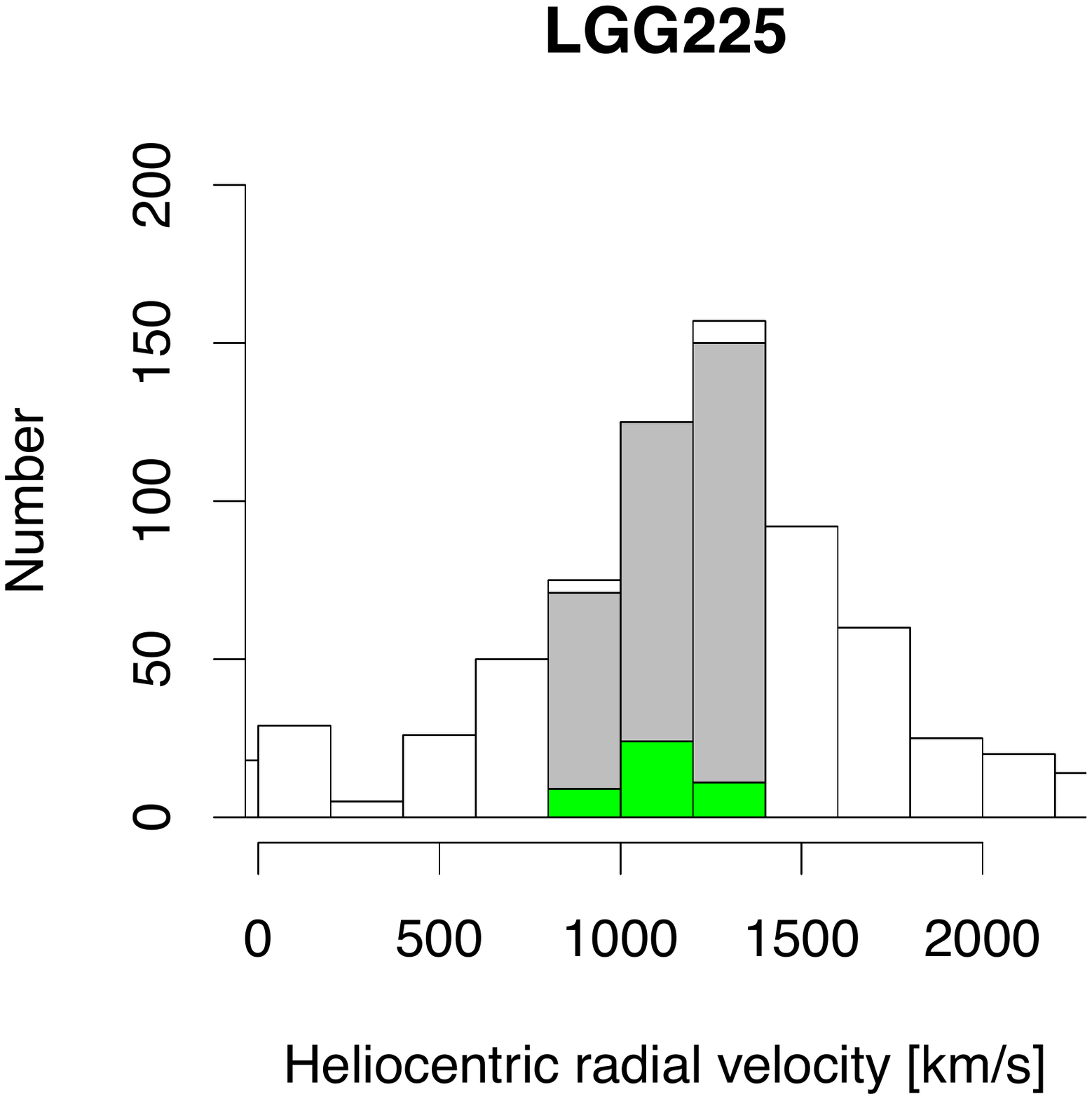}
\includegraphics[width=8cm]{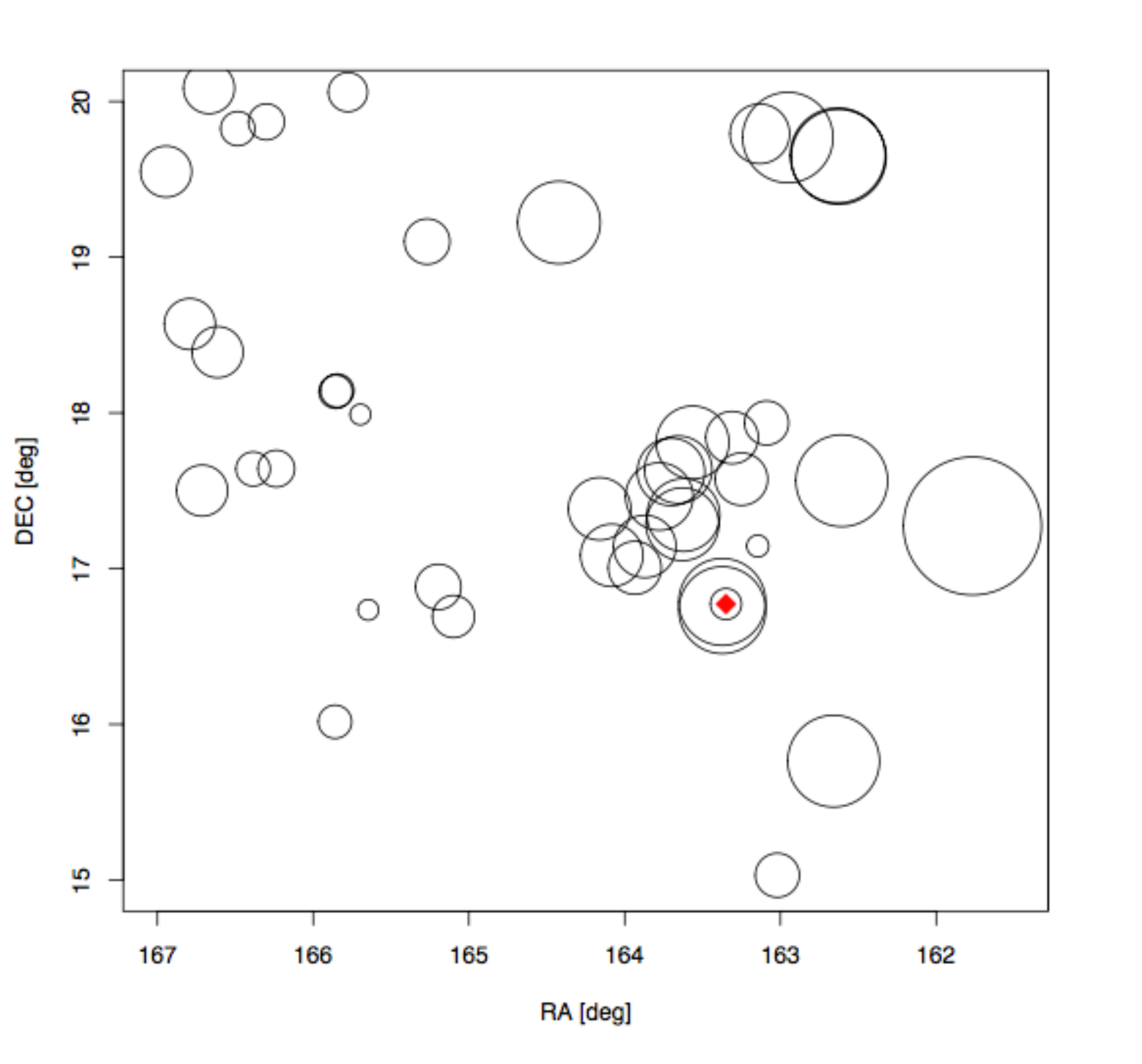}
\caption{
{\it Top left:} Spatial distribution of galaxies (black plus symbols) within a box of 4$\times$4\,Mpc$^2$ centred on LGG225)  \citep[and Table \ref{newdyn}]{Marino2010} and with  velocity extent, $\sigma$, defined by the dynamical analysis. % in \citet[their Table 6, B-band]{Marino2010}.
Blue square symbols show the members of the group in \citet{Marino2010},
green diamonds indicate the added members, and the big red filled diamond marks the position of NGC~3447/NGC~3447A.
The 2D binned kernel-smoothed number density contours of galaxies  fainter then  m${_B}$=15.5\, mag (crosses) are also shown  (red/yellow shaded area).
{\it Top right:}  Histogram of the heliocentric radial velocity of all the galaxies within the same box in the left panel %of 4$\times$4\,Mpc$^2$  
accounting for a velocity dispersion $\pm$1000\,km/s$^{-1}$ around the average velocity of the group (Table \ref{newdyn}). The width of the velocity bins is 200\,km\,s$^{-1}$. The filled (grey) bins  show the velocity distribution of galaxies with the same velocity extent as the members of the group;  the green bins
 account for the all the (old$+$new) members of the group, i.e. galaxies within a box of 1.5$\times$1.5\,Mpc$^2$ centred on LGG225 (Table \ref{LGG225}). 
{\it Bottom:} The \citet{DS88} "bubble-plot" based on the 10 nearest galaxies. The bubble size is proportional to the squared deviation of the local velocity distribution from the group velocity distribution. 
}
\label{figHistVh4Mpc}
\end{figure*}

NGC~3447/NGC~3447A belongs to LGG 225 group which occurs in a very crowded region of  the Leo Cloud (Fig.\ref{figHistVh4Mpc}, top-left panel). 
The population of Spirals, which dominate this group,
spans a range of ages between few Gyr to 6\,Gyr and their total mass ranges from 5 to 35$\times$10$^{10}$\,M$_{\odot}$ \citep{Marino2010}.
We revised the dynamical properties of this group following the recipes in
\citet{Marino2013} and \citet{Marino2016}. This allows us to improve the membership of the group increasing its number of galaxies  from 15 \citep[including NGC~3447A]{Marino2010} to 41 (Table \ref{LGG225}). 
Figure \ref{figHistVh4Mpc} (top right panel)  highlights the galaxy density and velocity distribution within a  box of 4$\times$4\,Mpc$^2$ centred on this group.
Table \ref{newdyn} points out the updated kinematical and dynamical properties of the group resulting from a non-luminosity weighted analysis of its members based on the recipes in \citet{Perea90} and \citet{Firth2006},
adopting H$_0$=75\,km\,s$^{-1}$\,Mpc$^{-1}$. As discussed  in these papers, a large scatter in the properties of the group could arise from this approach
if we weight by luminosity or not,
due to the different distributions of mass and light,
and on the adopted wavelength \citep{Marino2010}. Looking at Table \ref{newdyn}, we note a large difference between the virial  and projected mass estimates which points towards the presence of anisotropy in the group. This fact, together with the relatively long crossing time ($>$0.2 Hubble time), indicate that this group is not virialized \citep{Firth2006}. This conclusion is further  strengthened  by the "bubble-plot"  \citep{DS88} in Fig.\ref{figHistVh4Mpc} (bottom panel) showing a large number of small overlapping bubbles in the region of LGG 225. This means that  there are numerous  substructures lacking of any global potential field, i.e. a larger bubble including several/all of these substructures.
 Our simulations suggests that one encounter is enough to cause such a morphological, dynamical and photometric perturbation.
As a proxi of the role played by galaxy-galaxy interaction, we  inspect the group looking for both interacting (pairs) and 
perturbed members without an obvious companion. \citet{Marino2010} pointed out that LGG 225 group hosts a well 
known physical pair  NGC~3454/3455 (KPG 257, \citet{Karachentsev1972}).
 On the other side, there are several  galaxies showing morphological distortions/perturbation.  \citet{Marino2010} 
indicate UGC 6035 (the galaxy nearest to NGC~3447 in projection), 
UGC 6022, NGC 3370, NGC 3507 NGC 3501, UGC 6083, UGC 6112, UGC 6171. 
Taking into account that our simulation suggests that 
NGC~3447/3447A is a single, perturbed system, 
we conclude that the on-going interactions in the form of classical
physical pairs weight for about 5\% of the group members while about
22\% (9/41 including NGC~3447) is the fraction of galaxies with
un-ambiguous signatures of perturbations. Roughly 1 out 4 galaxies could
have been perturbed by encounters suggesting that the probability
of multiple interactions appears low. NGC~3447/3447A system with 
its unique morphology, according to our simulation, provides a snapshot of
the interaction job in this phase of the group evolution.

%%---------------------------- figure 9 
%-------------------------- end figure 9  ------------------------------------
%------------ table analisi dinamica---------------------
\begin{table*}
\center
\scriptsize
\caption{Kinematical and dynamic properties of the updated LGG225 group}
\begin{tabular}{llccccccccc}
\hline
Group& Center& V$_{group}$ & Velocity & D &Harmonic  & Virial  & Projected & Crossing&Group&\\
  name & of mass &       & dispersion &   & radius  &  mass &mass &time/H$_{0}$& Luminos.&\\
       &RA[deg]DEC&  [km/s$^{-1}$]& [km/s$^{-1}$]  & [Mpc] & [Mpc]  & [$10^{13}\,M_\odot$] & [$10^{13}\,M_\odot$]& &[$10^{11}\,L_\odot$] &\\
\hline\hline
LGG~225 &164.45676$+$17.88861  & 1116.84 &122.91& 14.89&  0.26&0.42&1.87& 0.35& 0.21  & \\
\hline
\end{tabular}
%\footnotesize{}
\label{newdyn}
\end{table*}
%----------end table analisi dinamica--------------------

\subsection{ The photometric properties of the group}
\begin{figure*}
\centering
\includegraphics[width=6.5cm]{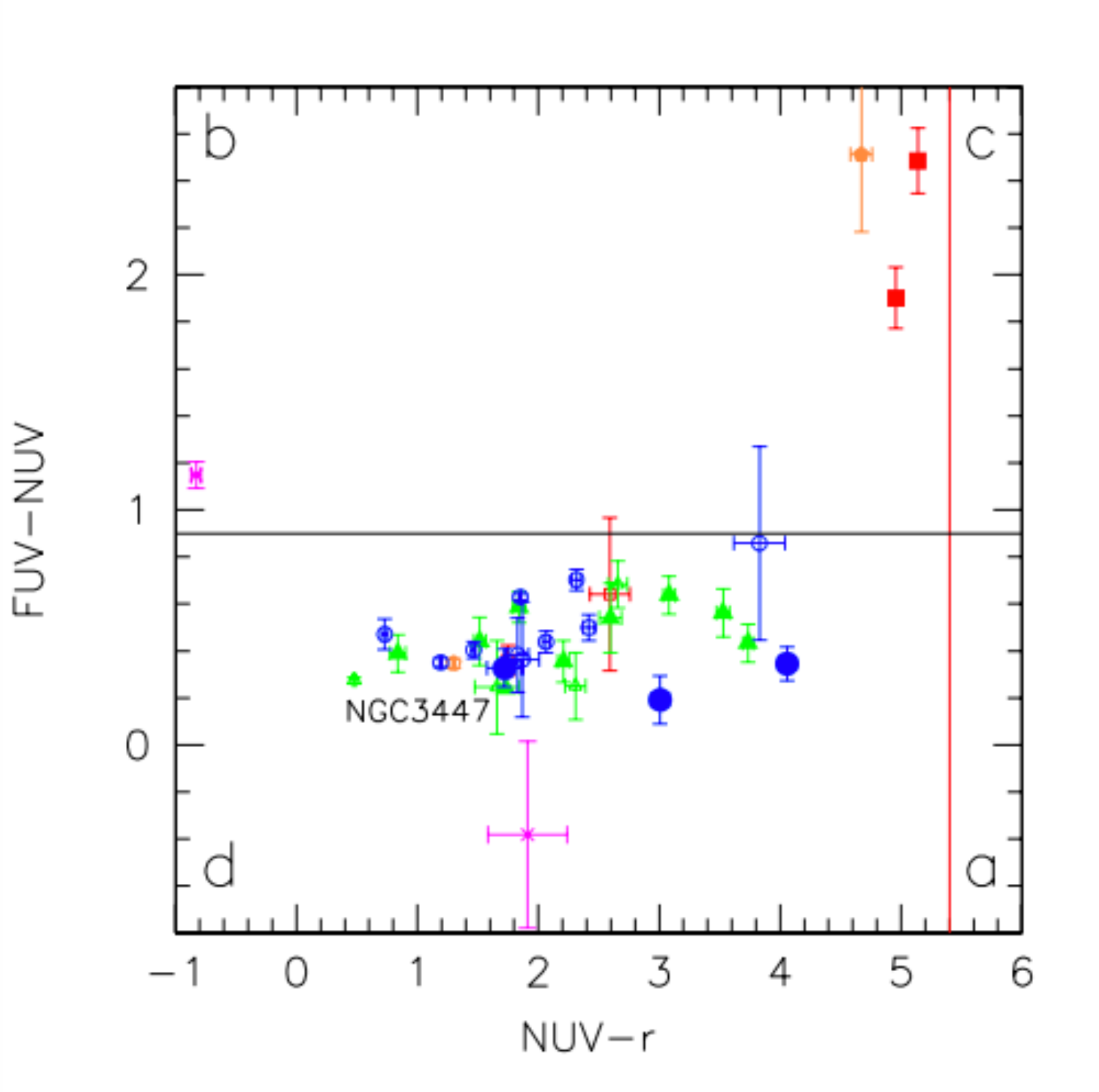}
\includegraphics[width=6.5cm]{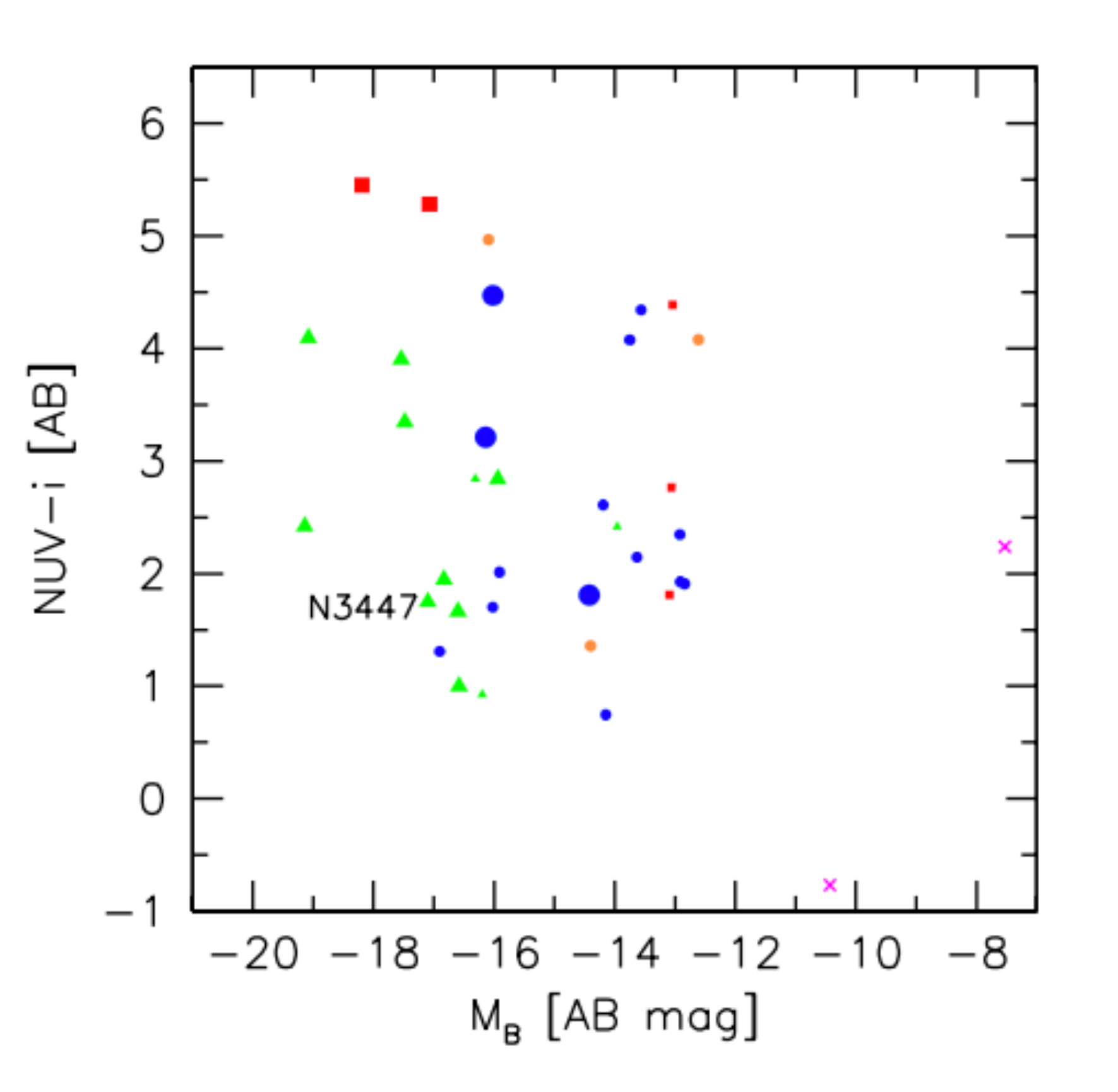}
\caption{
The color-color diagram,  NUV$-$r vs. FUV$-$NUV (left panel) and the 
color-magnitude diagram,  M$_{B}$ vs NUV$-$i ( right panel), for all the LGG~225 members;
(red) squares indicate ETGs (T$\le$-2), (orange) diamonds S0s (-2>T$\le$0), 
(green) triangles  Spirals (T$<$9),  (blue) circles show Irregulars (T$\ge$9) and magenta crosses are for galaxies without morphological classification in the Hyperleda catalog; small dots are for new members of this works (Table \ref{LGG225}).
Solid lines (left) correspond to the conditions FUV$-$NUV <0.9, i.e UV rising slope, and NUV$-$r > 5.4 shows the region of young massive star-free galaxies. These conditions, following
the UV classification scheme by \citet{Yi11}, separate passive evolving ETGs (region c) from star forming galaxies (region a), see text. 
In both panels magnitudes were corrected by galactic extinction following \citet{BH82}.}
\label{fig7}
\end{figure*}

  Figure \ref{fig7}, left panel, highlights that its galaxy population is dominated by star-forming galaxies and  its ETGs are not passively evolving.
%To clarify this point %we remember that the slope of the UV spectrum is related to the temperature of the stars emitting in the UV and that the condition: FUV$-$NUV$=$0.9 indicates a flat UV spectrum in the $\lambda$ vs F$_{\lambda}$ domain,
%whereas a bluer FUV$-$NUV corresponds to a steeper slope.
\citet{Yi11} established a colour criterion to classify
ETGs according to their UV spectral morphology based on
three color conditions. Passively evolving ETGs would have
NUV$-$r > 5.4, and FUV$-$r > 6.6. These last values indicate
the confidence limit of the red sequence in the color magnitude diagrams NUV$-$r vs M$_r$ and FUV$-$r  vs M$_r$ respectively. ETGs
showing UV upturn with no residual star formation have to
obey a further condition, FUV$-$NUV<0.9. \\
The early-types galaxies (ETGs) of this group, namely NGC~3457  and NGC~3522, do not live in the region of passively evolving ETGs, i.e. region (c), neither in that showing UV-upturn with no residual star formation, that is region (d) where UV is due to the old star populations. \citet{Mazzei2014} analysed  the  multi-$\lambda$ properties of NGC~3457 finding that these are well matched by an encounter, not a merger,  between two halos of equal mass and perpendicular spin, as in the case of NGC~3447/NGC~3447A, but with a total mass 10 times larger than that. NGC~3522 is likely due to a major merger (mass ratio 1:1) of total mass 4$\times 10^{11}$\,M$_{\odot}$. This E galaxy in the B-band is one magnitude fainter than NGC~3457. Its evolution in the color magnitude diagram, NUV-r vs M$_r$, shows frequent ''rejuvenation'' episodes which, of course, cannot change its optical properties but strongly modify the UV ones.  Such behavior occurs in low luminosity ETGs more than in the brighter ones \citep{Mazzei2014}. 
Moreover, Fig.\ref{fig7} (right panel) shows that there are no very bright galaxies in the group (i.e. no galaxies with M$_B$<-20 mag) and  ETGs are not the brightest ones emphasising that luminosity and mass distribution, in particular in the short wavelength range (UV and B bands), are different. This points against
the virialization of this  group.\\
%----------   figura 10 
%-------------------end figura 10---------------------------------
\indent
The $B$-band absolute magnitudes of Fig.\ref{fig7}, corrected by galactic extinction following \citet{BH82} as well as the colors,
account for the average group distance, 15\,Mpc  \citep[their Table 6]{Marino2010}. % and Table \ref{newdyn} of this work,  and the flux values in \citet[their Tables 1, 3 and  4 respectively for B, NUV and FUV, and r, i]{Marino2010} and Table \ref{fl_LGG225}. 
Dwarfs,  i.e. galaxies fainter than M$_{B}$=-16\,mag \citep{Ta94}, are numerous, as expected, in this still not virialized group. The complete analysis of this group deserves, of course, further investigation. We outlined here that  the NGC~3447/NGC~3447A system 
lives in a region where encounters are possible and 
 several late-type galaxies, fainter than that, are present, in agreement with the scenario accounted for by our SPH-CPI simulation.

%------------ table LGG225---------------------
\begin{table*}
\center
\scriptsize
\caption{Members of the LGG 225 group}
\begin{tabular}{llccccccccc}
\hline
Galaxy& RA & Dec & Morph &Hel. Vel.  & logD$_{25}$ & logr$_{25}$&PA& B$_T$& & \\
 name & (J2000) & (J2000) & type &   &   &   & & & &\\
       &[h]&  [deg]&  &[km/s$^{-1}$] & [arcmin] & [deg]  &[deg]& [mag]&  &\\
\hline\hline
NGC3370& 10.78446& 17.27366 & 5.1 &1281$\pm$  3& 1.39& 0.24& 146.5& 12.24 $\pm$ 0.25& \\
UGC05945& 10.84043 &17.56424 & 9.7& 1137$\pm$  4& 0.92& 0.26 & 94.3& 15.14$\pm$ 0.29&\\
UGC05947& 10.84175 &19.64388& 10.0& 1252$\pm$  3& 1.09& 0.24&  25.0& 14.92$\pm$ 0.08&\\
UGC05948 &10.84392& 15.76334 &10.0& 1120 $\pm$ 2& 1.04 &0.53 & 35.0 &17.00$\pm$ 0.50&\\
SDSSJ105148.64+194605.4 &10.86351& 19.76823& 10.0& 1374$\pm$   2& 0.56 &0.30 & 45.8 &18.34$\pm$  0.50&\\
SDSSJ105204.79+150149.6& 10.86799& 15.03043& -5.0&  828$\pm$  38& -& - &  5.8& -&\\
PGC032630& 10.87261& 17.93527& -0.9 &1304 $\pm$  4& 1.03& 0.23&  18.4 &15.12$\pm$  0.32\\
UGC05989& 10.87551& 19.79229&  9.9& 1138$\pm$  2& 1.16 &0.49& 124.9& 14.30$\pm$ 0.29&\\
SDSSJ105234.93+170841.8& 10.87637& 17.14494& 10.0& 1118 $\pm$ 2& 0.49 &0.12& 142.2& 18.25$\pm$ 0.50&\\
NGC3443& 10.88336& 17.57442&  6.6& 1109$\pm$   6& 1.25& 0.30& 145.7& 14.58$\pm$  0.36&\\
NGC~3447& 10.88999& 16.77242&  8.8& 1087$\pm$  9& 1.53 &0.24 & 14.3& 14.47$\pm$ 0.58&\\
NGC~3447A& 10.89158& 16.78601&  9.9& 1094$\pm$  7& 1.17 &0.28& 107.2& 15.35$\pm$ 1.10&\\
SDSSJ105329.58+164359.6& 10.89155& 16.73324& 10.0& 1074$\pm$  2& 0.62& 0.09& 174.3& 17.82$\pm$ 0.50&\\
UGC06022 &10.90427& 17.81020 & 9.7 & 974$\pm$  3& 1.03& 0.24&  11.1& 16.76$\pm$ 0.43&\\
NGC3454  &10.90820  &17.34398 &  5.4 & 1104$\pm$  6 & 1.38 & 0.66 & 116.2 & 13.71$\pm$ 0.10 &\\
NGC3455  &10.90863 & 17.28473 &  3.1  &1104$\pm$  2 & 1.36 & 0.20  & 69.7 & 14.30$\pm$ 0.27 &\\
AGC208602& 10.91039& 17.63500& -& 1093$\pm$  4& - & -& -& -&\\   
NGC3457  &10.91350 & 17.62117  &-4.4 & 1159$\pm$  5 & 0.99 & 0.02 &-  & 12.98$\pm$ 0.22 &\\
PGC1533359 & 10.91852 & 17.46265 & 10.0  &1167$\pm$  1 & 0.74 & 0.25  & 27.3 & 16.79 $\pm$0.50 &\\
UGC06035 & 10.92473  &17.14189 &  9.9 & 1073$\pm$  2  &0.99 & 0.19 & 177.0 & 15.15$\pm$ 0.32 &\\
PGC032843 & 10.92901  &17.00501 & 10.0 & 1138$\pm$  2  &0.80  &0.04 &-  & 15.26$\pm$ 0.39 &\\
SDSSJ105619.93+170505.9& 10.93887& 17.08501& 10.0&  961$\pm$  1& 0.45& 0.04& -& 18.25$\pm$ 0.50&\\
SDSSJ105638.66+172301.2 &10.94407& 17.38372& -5.0&  948$\pm$  1& 0.58& 0.29& 157.5& 18.07$\pm$ 0.50&\\
SDSSJ105741.23+191321.1 &10.96145& 19.22252& 10.0 &1006$\pm$ 12& -& -& 167.6 &17.60$\pm$ 0.50&\\
UGC06083& 11.00661& 16.69221&  4.1&  941$\pm$  2& 1.13& 0.92& 142.5& 15.19$\pm$ 0.29&\\
SDSSJ110047.14+165255.5& 11.01310& 16.88213& -5.0& 1136$\pm$  2 &0.55& 0.16& 101.3& 18.07$\pm$ 0.50&\\
UGC06095 &11.01789& 19.10012&  9.8& 1115$\pm$  2& 1.04& 0.43&   5.0& 16.66$\pm$ 0.49&\\
UGC06112 &11.04311& 16.73486 & 7.4& 1034$\pm$  3& 1.28& 0.52& 120.7 &14.53$\pm$ 0.35&\\
NGC3501 &11.04647 &17.98943 & 5.8& 1126$\pm$  4 &1.63 &0.82  &27.4& 13.60$\pm$ 0.22&\\
SDSSJ110306.42+200335.7 &11.05181& 20.06002& 10.0 &1284$\pm$  5& - &- & 72.5 &21.16$\pm$ 0.50&\\
NGC3507 &11.05705& 18.13599&  3.1&  976$\pm$   7& 1.47& 0.07&  95.3& 12.07$\pm$  0.51&\\
PGC087171 &11.05732 &16.01622 &10.0 &1230 $\pm$  3& 1.01 &0.43&  17.2 &16.93$\pm$  0.50&\\
SDSSJ110456.82+173830.5&11.08245& 17.64178& 10.0&  908$\pm$   4& 0.80 &0.33 &113.8 & 17.38$\pm$  0.50&\\
AGC219617& 11.08669 &19.86889& -& 1199$\pm$   6& -& -& -& -& \\
PGC033523  &11.09237 & 17.63961 &  4.5 & 1040$\pm$  1 & 1.19 & 0.40 &  23.8 & 14.82$\pm$ 0.33 &\\
UGC06151 & 11.09896 & 19.82534 &  8.7 & 1332$\pm$  2  &1.08 & 0.02 & - & 14.95$\pm$ 0.14 &\\
PGC1558217  &11.10758 & 18.38999 &  8.0  &1248$\pm$  2 & 0.61  &0.15  & 60.9 & 17.18$\pm$ 0.50 &\\
NGC3522  &11.11124 & 20.08561 & -4.9 & 1206$\pm$  3 & 1.07  &0.28 & 114.7 & 14.07$\pm$ 0.29 &\\
SDSSJ110651.09+173002.8 & 11.11418  &17.50090 & -2.0 &  949$\pm$ 71 & 0.48 & 0.12  &153.8  &18.52$\pm$ 0.50 &\\
UGC06171 & 11.11949 & 18.57027  & 9.9 & 1203$\pm$  1 & 1.24 & 0.53 &  65.4 & 15.00$\pm$ 0.28 &\\
UGC06181 & 11.12962  &19.54938 &  9.7 & 1170$\pm$  3 & 1.02  &0.18 &  48.0 & 15.45$\pm$ 0.32 &\\
\hline
\end{tabular}
%\footnotesize{}
\label{LGG225}
\end{table*}

\end{appendix}
\end{document}